\documentclass{article}

\usepackage{PRIMEarxiv}

\usepackage[utf8]{inputenc} 
\usepackage[T1]{fontenc}    
\usepackage{hyperref}       
\usepackage{url}            
\usepackage{booktabs}       
\usepackage{amsfonts}       
\usepackage{nicefrac}       
\usepackage{microtype}      
\usepackage{lipsum}
\usepackage{fancyhdr}       
\usepackage{graphicx}       
\usepackage{multirow}
\usepackage{multicol}
\usepackage{nomencl}
\usepackage{etoolbox}
\usepackage{subfigure}
\usepackage{amsmath}
\usepackage{placeins}
\graphicspath{{media/}}     
\makenomenclature
\title{Toward Development of an Improved Friction Correlation for the Near-Wall Region of Pebble Bed Systems
\thanks{\textit{\underline{Citation}}: 
\textbf{Authors. Title. Pages.... DOI:000000/11111.}} 
}

\author{
  David Reger, Elia Merzari \\
  Pennsylvania State University \\
  University Park, PA 16803\\
  \texttt{dzr5281@psu.edu ebm5351@psu.edu} \\
   \And
  Paolo Balestra, Sebastian Schunert \\
  Idaho National Laboratory \\
  Idaho Falls, ID 83415\\
  \texttt{paolo.balestra@inl.gov sebastian.schunert@inl.gov} \\
  \And
  Yassin Hassan \\
  Texas A\&M University  \\
  College Station, TX 77843 \\
  \texttt{y-hassan@tamu.edu}
  \And
  Haomin Yuan \\
  Argonne National Laboratory \\
  Lemont, IL 60439 \\
  \texttt{hyuan@anl.gov}
}

\begin{document}
\maketitle

\begin{abstract}
The development of High-Temperature Gas-Cooled Reactors (HTGRs) and Fluoride-Cooled High-Temperature Reactors (FHRs) that utilize pebble fuel has drastically increased the demand for improving the capabilities to simulate the packed beds found in these reactors. The complex flow fields found in a pebble bed make computational fluid dynamics (CFD) simulations time consuming and costly. Intermediate fidelity porous media models, however, are capable of approximating these flow fields in a much more computationally efficient manner. These models require the use of closures to capture the effects of complex flow phenomena without modeling them explicitly. This research employs data obtained from high-fidelity CFD simulations of a pebble bed to improve the drag closures used in porous media models in the near-wall region of the bed. Specifically, NekRS, a GPU-enabled spectral element CFD code, was used to simulate a bed of 1,568 pebbles at Reynolds numbers of 2,500, 5,000, and 10,000. The case was divided into five concentric subdomains to extract radial profiles of the average porosity, velocity, and wall shear in each subdomain. A model consistent with the high-fidelity model was created in Idaho National Laboratory's Pronghorn porous media code and the KTA correlation was chosen as the drag closure of comparison. Averages for the velocity, friction factor, and form factor were extracted in each region of the Pronghorn results to provide the comparative data. It was found that the KTA correlation significantly overestimates the velocity in the near-wall region. An investigation of the drag coefficients between the two codes revealed that the KTA correlation underestimated the form factor in the outermost region while overestimating it in the inner four regions. Manually increasing this form factor in the Pronghorn case predicted a velocity distribution in better agreement with the NekRS result.

An investigation of the pressure drop was also performed, in which the NekRS results were compared to the KTA correlation, the manually adjusted Pronghorn values, and the Wentz/Thodos correlation. It was found that the KTA correlation overpredicted the pressure drop at all three Reynolds numbers. The Wentz/Thodos correlation performed similarly to the KTA correlation, although it underpredicted the pressure drop in Re = 5,000 and Re = 10,000 cases. Although the manually increased drag coefficients saw better velocity agreement with the NekRS result, the increase in the form factor worsened the overprediction of the pressure drop.

This analysis in this work has revealed the underlying inaccuracy in the near-wall region of the KTA correlation and has set up the process for using high-fidelity simulation to predict more accurate drag coefficients a priori, rather than with a manual velocity-matching approach. This process will allow for the devlopment of an improved drag closure for use in porous media models.
\end{abstract}


\section{Introduction}
Packed beds are an integral part of emerging generation-IV nuclear reactors. Some of these reactors plan to utilize pebble fuel in which the fissile material is embedded into spheres of graphite. These graphite spheres are loaded into the reactor vessel, creating a randomly packed bed of spheres. The coolant in these designs, typically either an inert gas or a molten salt, flows through the packed bed, creating a complex flow field. Understanding the intricacies of this complex flow is critical to advancing the deployment of pebble bed reactors. 

This work focuses in particular on an investigation of the wall-channel effect in packed beds. This effect is found in the near-wall region of a packed bed where the packing of the pebbles is disrupted by the presence of the wall. This disruption causes them to pack less efficiently, creating more space between pebbles and thus increasing the porosity near the wall. The increased porosity has a number of implications on the flow velocity, pressure drop, and heat transfer in this region and thus must be understood in order to accurately model Pebble Bed Reactors (PBRs).

Accurately modeling the wall-channel effect in a packed bed requires an advanced understanding of the pressure drop in each region of the system. Many correlations have been previously developed to predict the pressure drop. These correlations typically represent the pressure loss as the sum of viscous (Darcy) and inertial (Forccheimer) energy losses. One of the first major correlations to predict the pressure drop in packed beds in this manner was developed by Ergun \cite{Ergun}. Mehta and Hawley \cite{MehtaHawley} later found that Ergun's correlation saw significant discrepancy from experimental results for $d_{bed}/d_{peb} < 50$ as a result of the larger role that wall effects play in low bed-to-pebble diameter ratio cases. They suggested improvements to Ergun's correlation by redefining the equivalent diameter to include the size of the bed to account for wall effects. This correlation was later improved by Riechelt \cite{Reichelt} through the use of experimental studies to account for a larger range of Reynolds numbers, porosities, and bed-to-pebble diameter ratios. Hicks \cite{Hicks} also provided improvements to the Ergun equation by suggesting that the constants in front of the viscous and inertial terms are not constant, but rather depend on the Reynolds number. More recently, Choi \cite{Choi} developed a model based on the Ergun equation for small bed-to-pebble diameter rations. This correlation included a wall correction factor for the inertial loss term. Wu et. al. \cite{Wu} performed a study to investigate the effect of bed height on the pressure drop and developed a correlation to reflect this effect. The German Nuclear Safety Standards Comission (KTA) developed a correlation for specific application to High Temperature Gas-Cooled Reactors (HTGRs) \cite{KTA}. This correlation is the main comparative correlation for this study for this reason. Wentz and Thodos have also developed a correlation for a wide range of Reynolds numbers and porosities \cite{WentzThodos}. This wide range of validity makes this correlation valuable for comparison in this study as well.

Other studies similar to this work have also been conducted to compare CFD data to available correlations. Yildiz et. al. \cite{Yildiz} have performed DNS simulations to compare the results to multiple of the previously mentioned correlations. Atmakidis and Kenig \cite{Kenig} have also performed a study analyzing the wall effects in regular and irregular packed beds to compare to available correlations. Jun et. al. \cite{Jun}have simulated bypass flow rates in the HTR-PM HTGR. 

\begin{figure}
    \centering
    \includegraphics[width=0.75\textwidth]{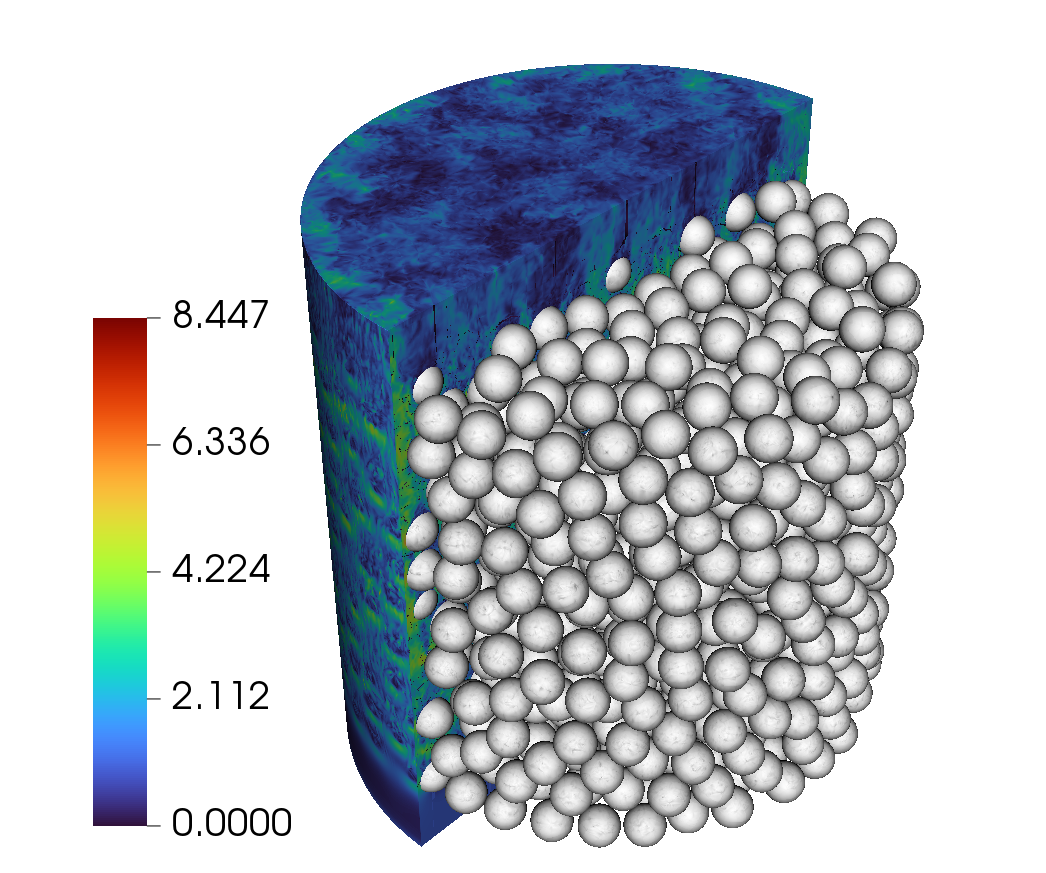}
    \caption{1,568 pebbles case from the previous work in this line of study with instantaneous velocity field shown on the left side.}
    \label{fig:ICONEFig}
\end{figure}

Previous work in this line of study focused on comparing the velocity distributions between high-fidelity LES simulation and intermediate-fidelity porous media simulation \cite{RegerICONE}. In this work, the KTA correlation was used as the drag coefficient in Idaho National Lab's Pronghorn porous media code. Two cases of 1,568 pebbles and 45,000 pebbles were created and simulated in NekRS at a Reynolds number of 20,000. A visualization of the 1,568 pebbles case with the velocity field can be found in Figure~\ref{fig:ICONEFig}.

The cases were separated into five concentric subdomains in postprocessing to extract porosity and velocity averages as a function of the radial position. The four outermost rings had a thickness of 0.25$d_{peb}$ each and the inner ring served contained the rest of the bed. The use of five rings was chosen to provide sufficient resolution of the near-wall region while also allowing for the combining of rings in the future if a coarser description of the wall region is desired. Two cases consistent with the NekRS cases were created in Pronghorn. It was found that the KTA correlation was causing the velocity in the near-wall region to be vastly overestimated and the distance needed to redistribute amongst the five subdomains was significantly longer than what was seen in the NekRS case. Manual changes to the drag coefficients in the near-wall region demonstrated significantly better agreement with the NekRS results, and a reduction in the non-flow direction drag coefficients improved the flow redistribution distance as well. This work demonstrated that capturing the wall-channeling effect in the porosity alone of a porous media model is not sufficient to reproduce the effect on the velocity field. It was determined that an improved correlation would be necessary to improve the ability to model the drag in the outer region. In this previous work, pebble-to-pebble contact points were handled by slightly shrinking the diameter of each pebble such that the pebbles do not touch in the resulting mesh. This approach, however, creates pressure gradients that are significantly lower than what is expected from a packed spherical bed. The major change to the mesh for this work is that the pebble contact is now included in the mesh, and a small chamfer is placed at the point of contact to prevent singular mesh points from appearing.

This work aims to continue the investigation of porous media codes' ability to model the wall-channel effect by examining the pressure drop in high-fidelity and intermediate-fidelity simulations. An investigation of the pressure drop provides additional insight into the ratio of friction and form losses in the different regions of a packed bed and will identify the areas of discrepancy that are seen with available correlations. Large-scale simulation with the Nek5000/NekRS spectral element code provide the benchmark data, and the KTA and Wentz/Thodos correlations are used are the pressure drop correlations of choice.

\section{Methods}
\subsection{NekRS}
The high-fidelity code chosen for this work is Argonne National Laboratory's NekRS spectral element code. This code is a relatively new GPU-oriented version of the established open-source spectral element code Nek5000 \cite{fischer2016}. NekRS is written in C++ and it is able to link to Nek5000 to leverage the existing extensive pre- and post-processing utilities. It was created as part of the DOE's Exascale Computing Project. NekRS is able to realize high throughput on advanced GPU nodes and demonstrates excellent scalability, and further details of NekRS performance for nuclear application have been provided in a recent publication \cite{merzari2020}.

Vadlidation of the Nek5000/NekRS code has previously been performed in a number of existing works. Specifically for pebble beds, work by Yildiz et. al \cite{Yildiz} has demonstrated good agreement of first-order statistics between Nek5000 and experimental results. This work also demonstrated that the meshing method of applying chamfers at pebble contact areas does not significantly influence the porosity of the resulting bed. Additional verification and validation studies for Nek5000/NekRS can be found in works by Lai et. al \cite{Lai} and Obabko et. al \cite{Obabko}.

Calculations performed as part of this work are wall-resolved Large Eddy Simulations (LES). A high pass filter approach is used for to mimic the effect of dissipation by the subgrid scales \cite{2021nek5000}. 

The mesh for this study was created using a novel voronoi cell approach developed as part of the Cardinal multi-physics project \cite{Cardinal}. First, pebble centers were generated by performing a Discrete Element Method (DEM) calculation with a target bed porosity of 40\%. This method then generates an all-hexahedral mesh for the pebble void regions based on a tessellation of the voronoi cells that are based on the previously generated pebble centers. This method assumes that all pebbles are initially perfectly spherical, then a small chamfer is placed in locations where pebbles contact each other or the wall. The resulting mesh contains 1,568 pebbles at an average porosity of roughly 44\%. A cross section of the mesh can be found in Figure~\ref{fig:meshes}. Further information on the meshing strategy can be found in \cite{LanMeshing}.

The model was simulated using NekRS on the Summit supercomputer across 228 GPU's. The case was initally simulated at Reynolds numbers of 2,500, 5,000, and 10,000 based on the pebble diameter. Cases were run until the ring-average velocity results converged to a constant value.

\subsection{Postprocessing of CFD data} \label{sec:postpro}
A key aspect of this work has been the development of a post-processing framework to reduce the CFD data to a form usable in comparisons with porous media models.
Postprocessing was  performed on the time-averaged fields to extract porosity, velocity, pressure, and wall-shear data from five concentric regions. These quantities were extracted from an axial slice of height $2.5 d_{peb}$ in the center of the bed to avoid the influence of entrance or exit effects.

In order to determine the friction and form coefficients from this data, the following process was used. First, the wall shear stress was used to determine the friction factor using the following formula:

\begin{equation}
    f = \frac{{8}\tau_{wall}}{\rho{v_s}^2}
\end{equation}

Then, the pressure gradient from frictional losses was determined with the Darcy-Weisbach equation.

\begin{equation}
    \frac{\Delta{P}}{L}_{friction}= f\left(\frac{\rho}{2}\right)\left(\frac{{v_s}^2}{D_h}\right) \qquad \textnormal{where} \quad D_h = \frac{\epsilon{d_{peb}}}{1-\epsilon}
\end{equation}

The pressure gradient from form losses was then calculated by subtracting the frictional gradient from the total pressure gradient

\begin{equation}
    \frac{\Delta{P}}{L}_{form}=\frac{\Delta{P}}{L}_{total}-\frac{\Delta{P}}{L}_{friction}
\end{equation}

Finally, the form coefficient was determined from the form pressure gradient.

\begin{equation}
    C_{form} = \frac{\frac{\Delta{P}}{L}_{form}}{0.5\rho{v_s}^2}
\end{equation}

\subsection{Pronghorn}

Idaho National Laboratory's Pronghorn porous-media code was used as the intermediate-fidelity code for this work. Pronghorn is a finite element thermal-hydraulics code built on the Multiphysics Object Oriented Simulation Environment (MOOSE) framework. It is intended to provide steady-state and transient simulation results in short code execution times to assist in design scoping studies or to provide boundary conditions for system-level analysis. Pronghorn is capable of multiple solver models, but the compressible porous media Navier-Stokes model was chosen for this work. The strong form of this model is given in \cite{PronghornTheoryManual}.

\begin{equation} \label{eqn1}
  \epsilon\frac{\partial{\rho}}{\partial{t}}+\nabla\cdot(\epsilon\rho_{f}\vec{V})=0,
\end{equation}
\begin{equation}
 \begin{aligned}
  \epsilon\frac{\partial{(\rho_{f}\vec{V})}}{\partial{t}}+\nabla\cdot(\epsilon\rho_{f}\vec{V}\vec{V})+\epsilon\nabla{P} -\epsilon\rho_{f}\vec{g}+W\rho_{f}\vec{V} -\nabla\cdot(\mu\nabla\vec{V})=0,
 \end{aligned}
\end{equation}
\begin{equation}
 \begin{aligned}
  \epsilon\frac{\partial{(\rho_{f}{E}_{f})}}{\partial{t}}+\nabla\cdot(\epsilon{H}_{f}\rho_{f}\vec{V})-\nabla\cdot(\kappa_{f}\nabla T_{f})-\epsilon\rho_{f}\vec{g}\cdot\vec{V} +\alpha(T_{f}-T_{s})-\dot{q_{f}} = 0,
 \end{aligned}
\end{equation}
\begin{equation}
  (1-\epsilon)\rho_{s}{C}_{p,s}\frac{\partial{T_{s}}}{\partial{t}}-\nabla\cdot(\kappa_{s}\nabla{T}_{s})+\alpha(T_{s}-T_{f})-\dot{q_{s}} = 0.
\end{equation}

\begin{figure}
    \centering
    \includegraphics[width=0.5\textwidth]{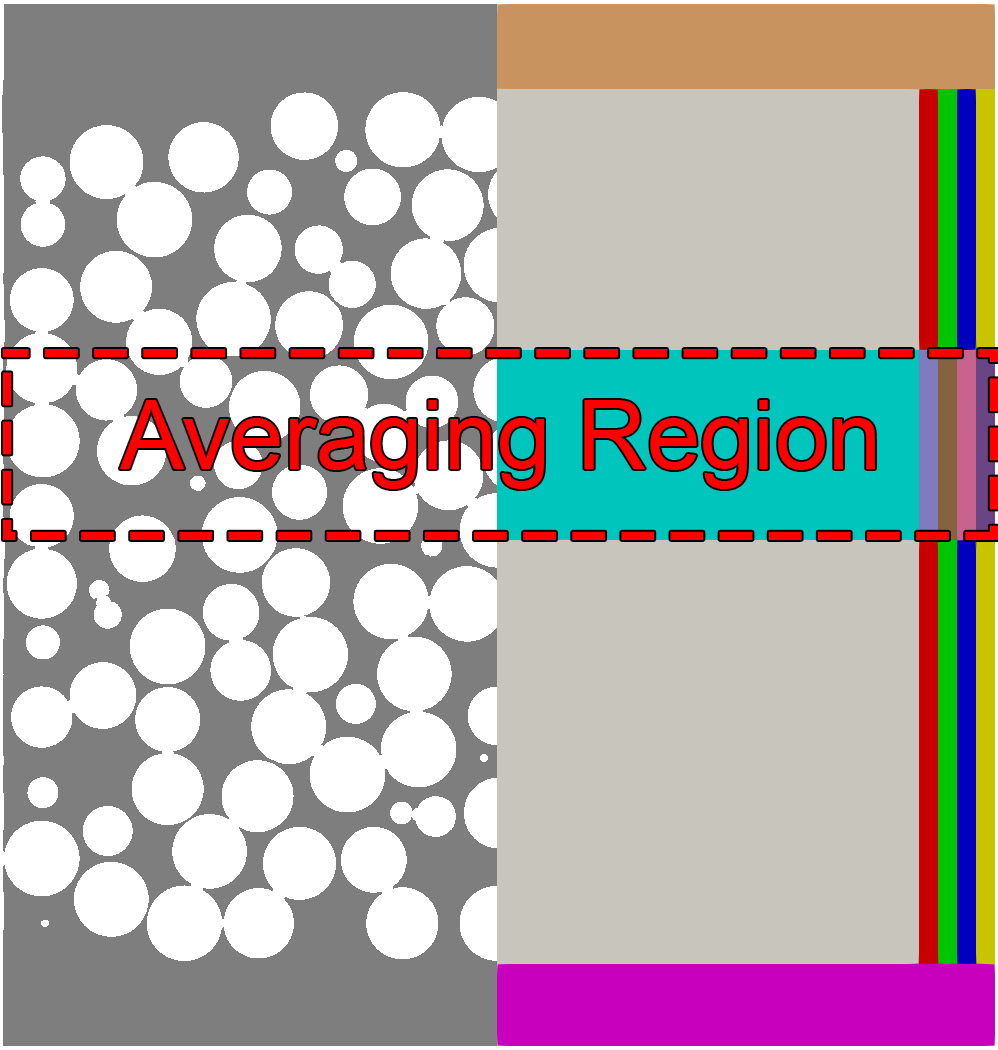}
    \caption{Centerplane slice of the NekRS mesh with the Pronghorn mesh overlayed on the right side. The averaging region where values were extracted from in both cases is indicated with the red dashed line.}
    \label{fig:meshes}
\end{figure}

A model consistent with the NekRS model was created in Pronghorn to provide comparative data. A nondimensionally identical RZ mesh was used to create these models. This mesh included the five subdomains where the porosity values extracted from the NekRS model were explicitly assigned. The Pronghorn mesh is pictured in Figure~\ref{fig:meshes} overlayed over a centerplane slice of the NekRS model. Helium was selected as the coolant fluid, and the inlet velocity was scaled to achieve each of the three average Reynolds numbers of the NekRS cases. Averaging was performed in the center of the model in the same region as the NekRS model. The average velocity, pressure, friction coefficient, and form coefficient were extracted for each of the five subdomains.

\FloatBarrier
\subsection{KTA Drag Correlation}

The KTA drag correlation developed by the German Nuclear Safety Standards Commission was the primary correlation that was investigated for this study. This correlation was developed specifically for application to HTGR's and is valid only over a narrow range of porosities from $0.36 < \epsilon < 0.42$. It should be noted that the bed used for this study had an average porosity slightly outside of this range.  The KTA correlation also does not account for any wall-channeling effects. The range of valid Reynolds numbers, however, is quite large at $10 < {Re}_{m} < 100,000$. The correlation can be found in Equation~\ref{eq::KTA}.

\begin{equation}
    \frac{\Delta{P}}{L} = \left(\frac{320}{{Re}_{m}}+\frac{6}{{{Re}_{m}}^{0.1}}\right)\left(\frac{1-\epsilon}{\epsilon^3}\right)\left(\frac{\rho^2{v_{s}}^2}{{d}_p}\right)\left(\frac{1}{2\rho}\right)
    \label{eq::KTA}
\end{equation}

\subsection{Wentz and Thodos Drag Correlation}

The second drag correlation used to compare in this work was the Wentz and Thodos correlation. This correlation was selected due to the large range of valid porosities compared to other correlation, with the range quoted as $0.354 < \epsilon < 0.882$, meaning that the simulation case is in the range of the correlation. This correlation as well does not explicitly account for any wall-channeling effects. The range of valid Reynolds number is smaller than the KTA correlation, but the cases used for this work are within the range of $2,540 < Re_{m} < 64,900$. It was developed using pebbles of 3.1242 cm diameter and a bed-to-pebble diameter ratio of 11.382. The correlation can be found in Equation~\ref{eq::WentzThodos}.

\begin{equation}
    \frac{\Delta{P}}{L} = \left(\frac{\mu{v_s}}{{d_{p}}^2}\right)\left(\frac{(1-\epsilon)^2}{\epsilon^3}\right)\left(\frac{0.396{Re}_{m}}{{{Re}_{m}}^{0.05}-1.20}\right)
    \label{eq::WentzThodos}
\end{equation}

\FloatBarrier

\FloatBarrier
\section{Analysis and Results}

\begin{figure}
    \centering
    \subfigure{
    \includegraphics[width=0.31\textwidth]{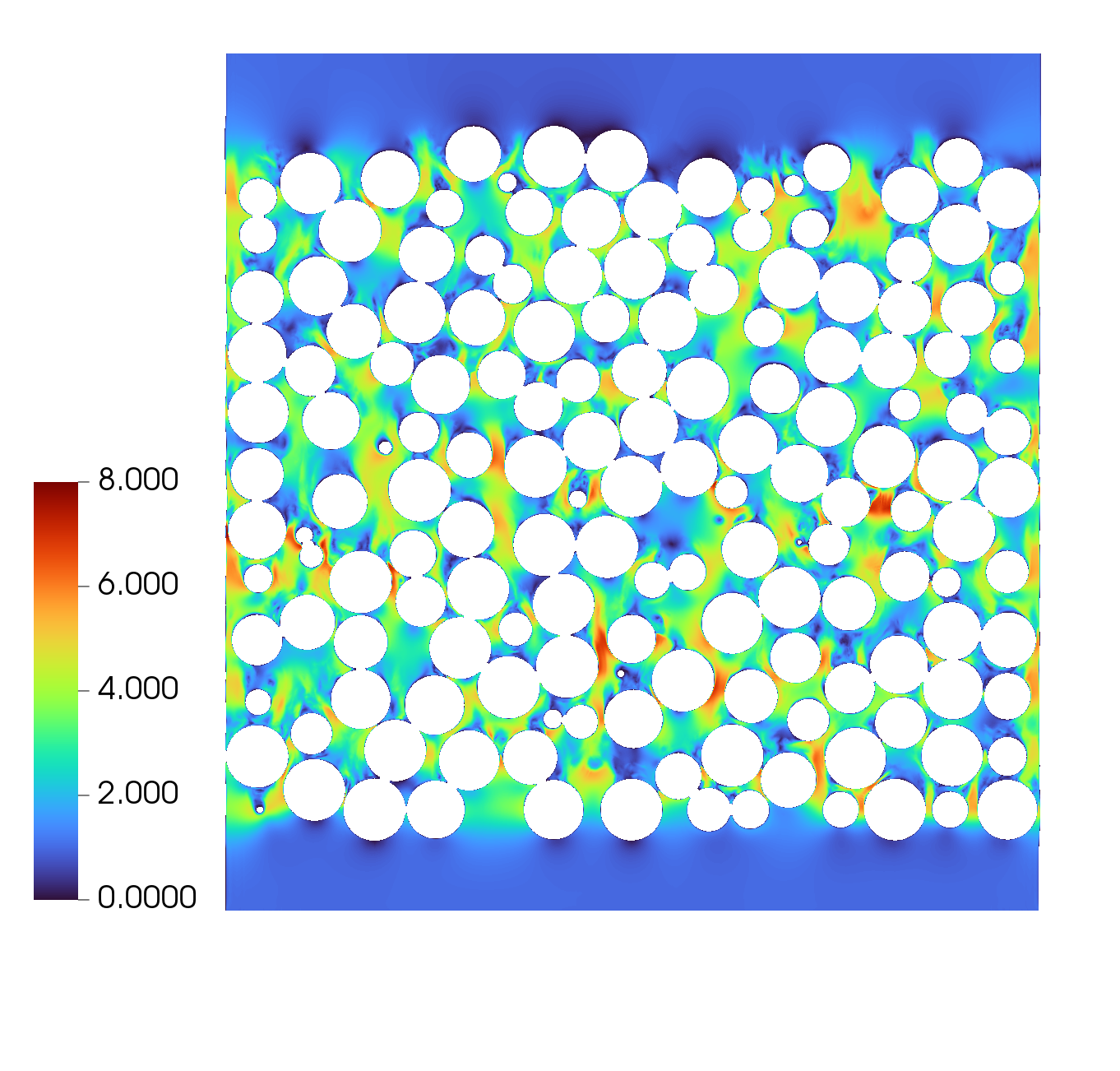}
    }
    \subfigure{
    \includegraphics[width=0.31\textwidth]{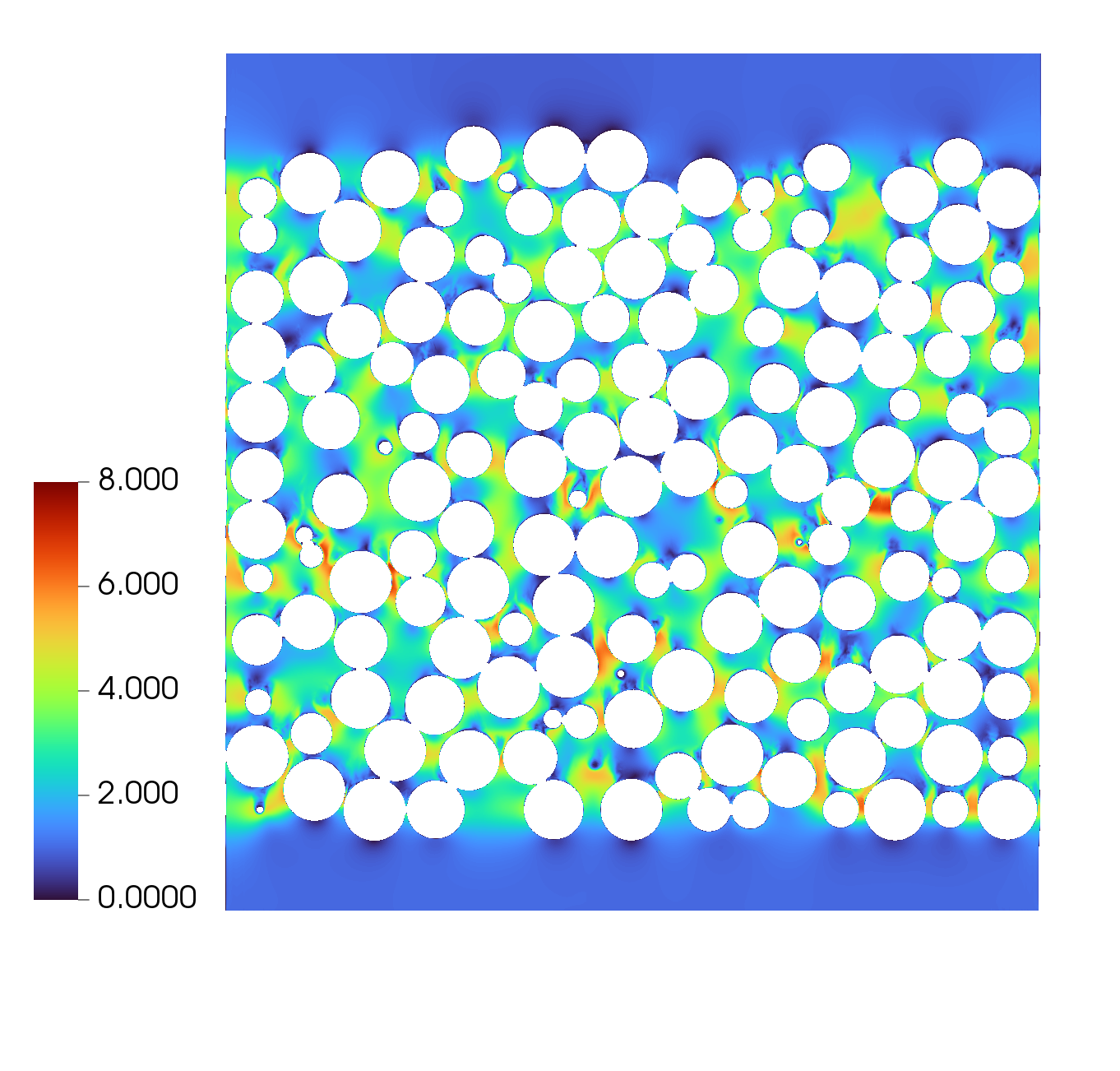}
    }
    \subfigure{
    \includegraphics[width=0.31\textwidth]{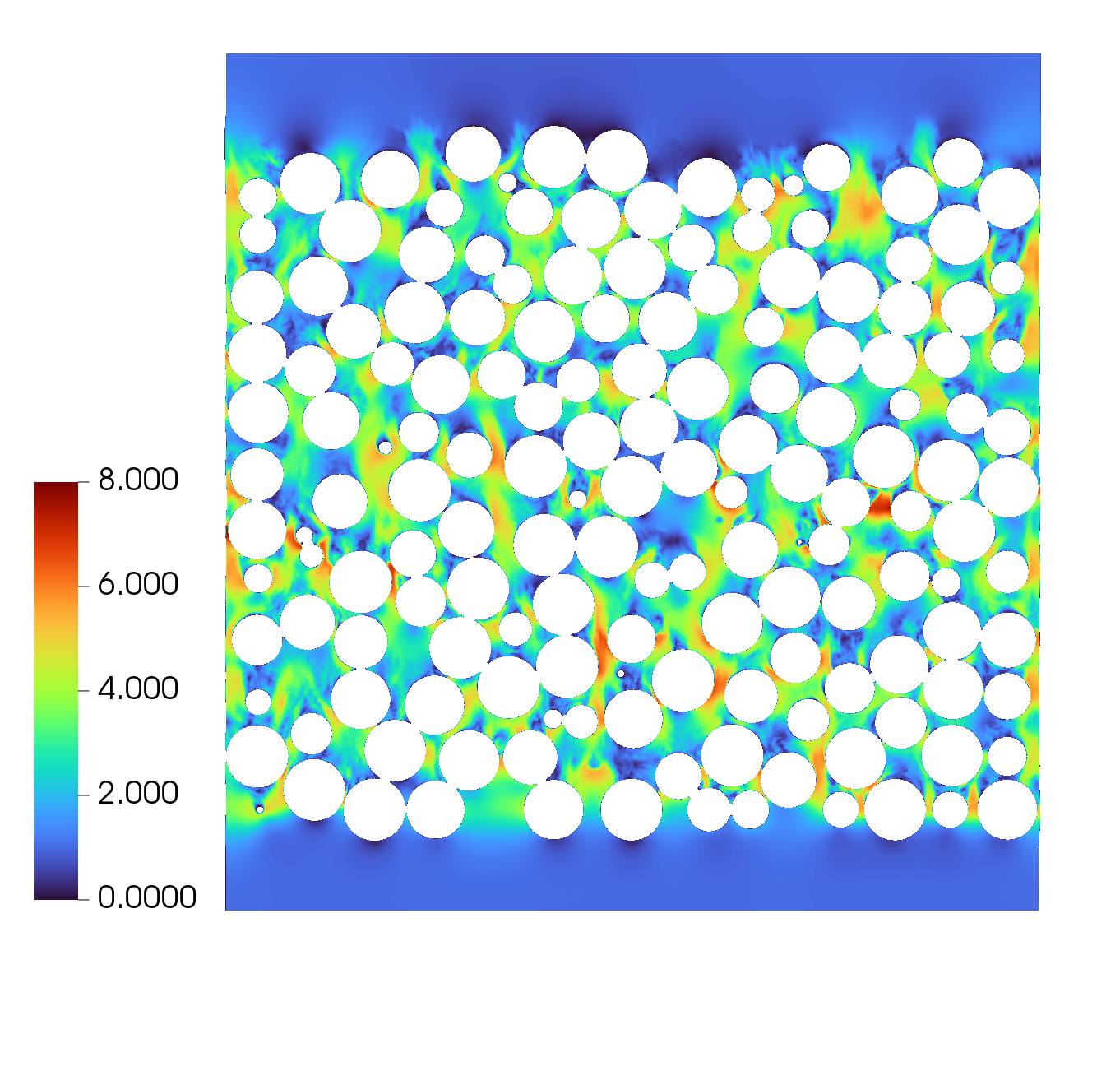}
    }
    
    \caption{Time-averaged velocity fields at the centerplane of the 1,568 pebbles cases. Re = 2,500 (left), Re = 5,000 (center), Re = 10,000 (right)}
    \label{fig:1568AvgFields}
\end{figure}

\begin{figure}
\centering
    \subfigure{
    \includegraphics[width=0.31\textwidth]{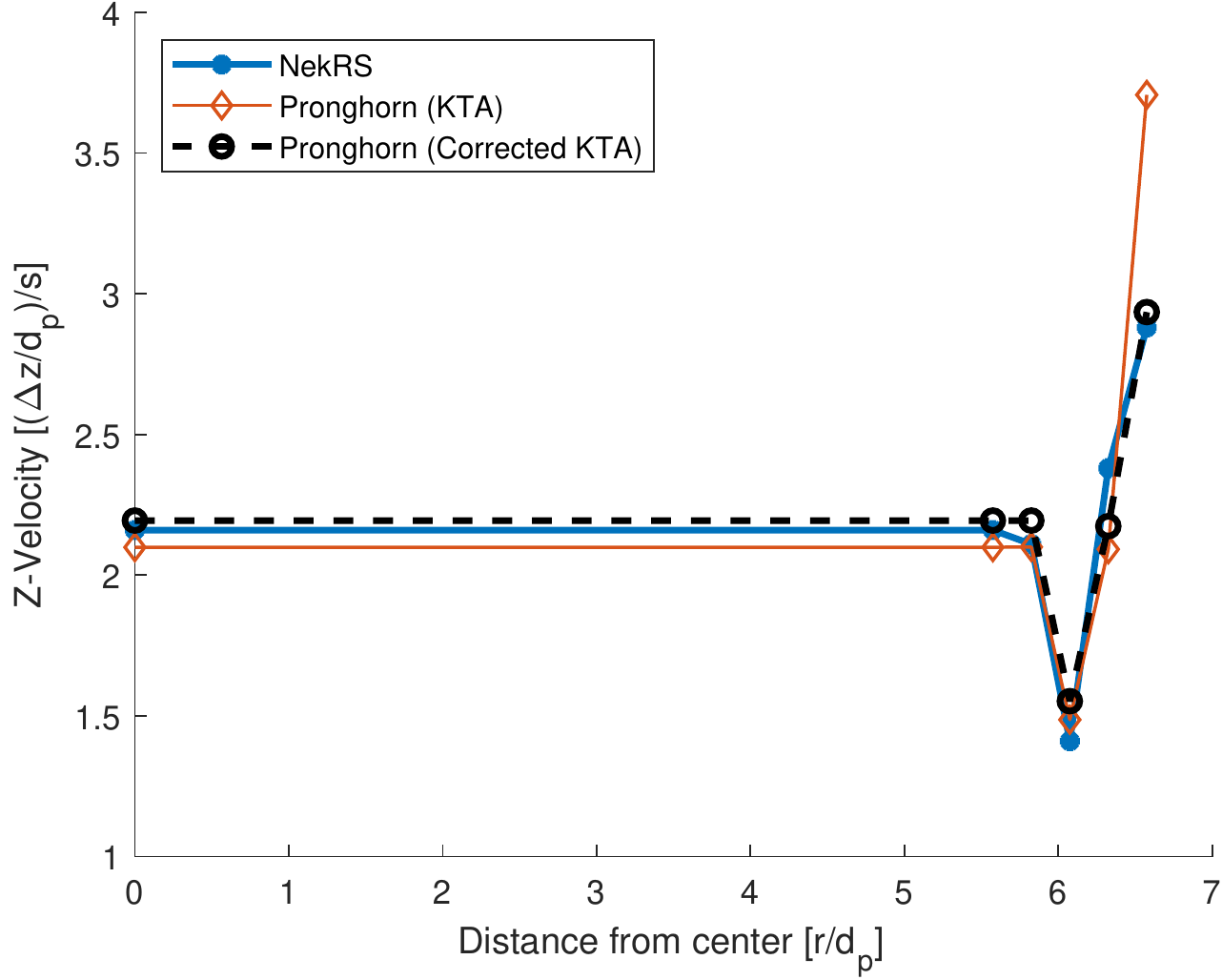}
    }
    \subfigure{
    \includegraphics[width=0.31\textwidth]{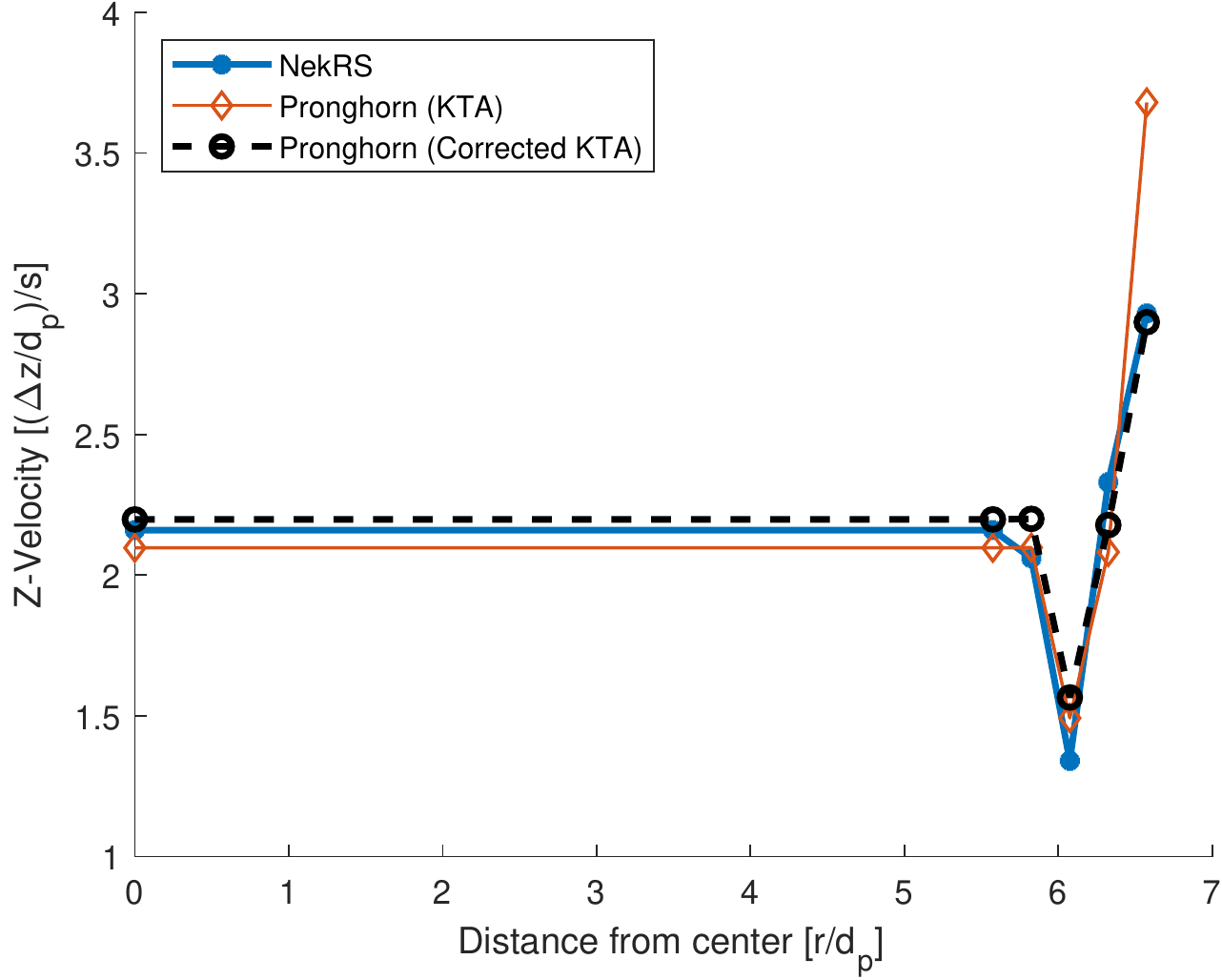}
    }
    \subfigure{
    \includegraphics[width=0.31\textwidth]{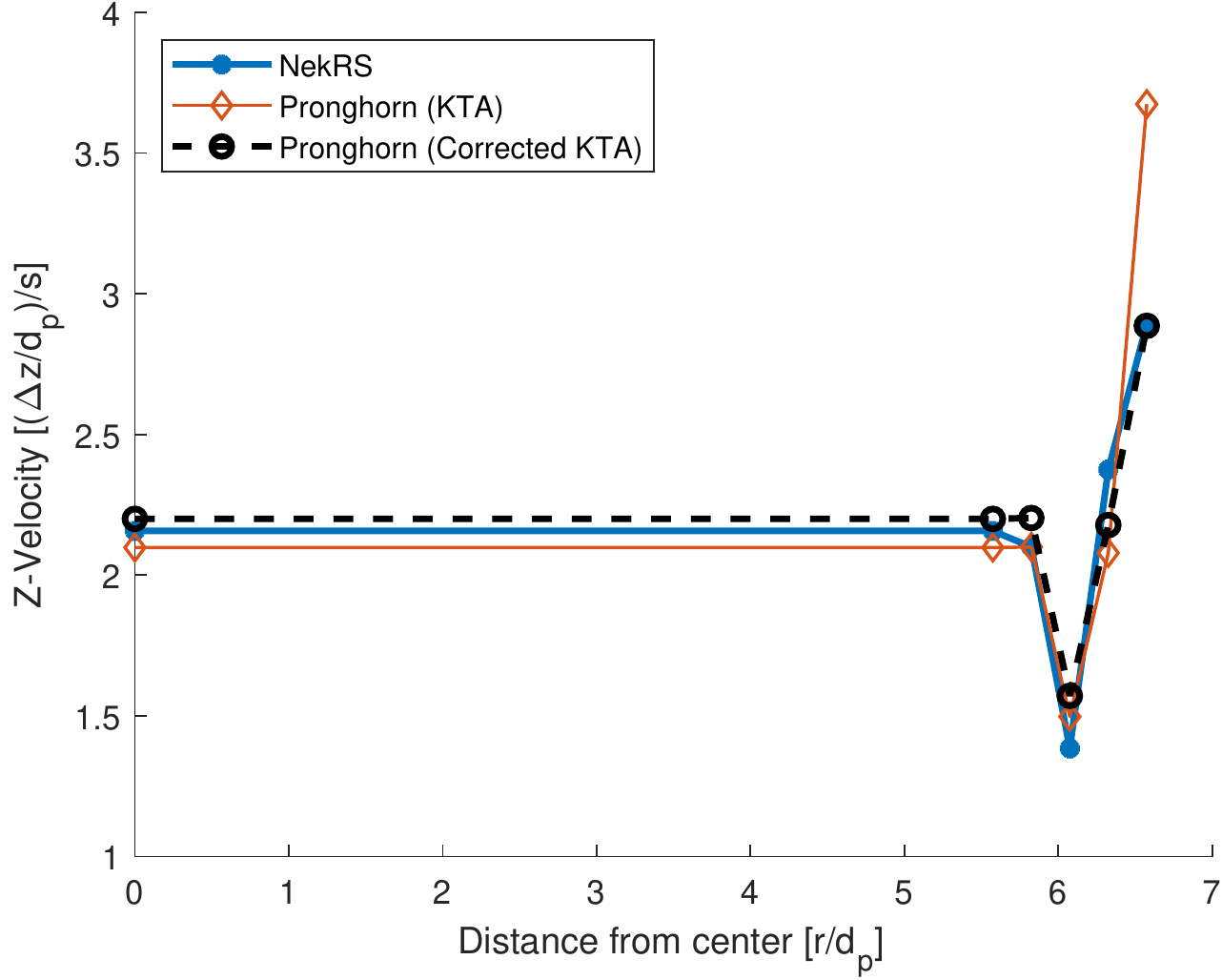}
    }
    
    \caption{Radial velocity profiles for the NekRS, Pronghorn KTA, and Pronghorn with manual adjustments cases}
    \label{fig:velprofiles}
\end{figure}

The pebble bed was simulated with NekRS at the three specified Reynolds numbers and the averaged fields can be found in Figure~\ref{fig:1568AvgFields}. The wall-channeling effect is visible in these images, as the velocity is noticeably higher in the near-wall region. From these average flow fields, the porosity, velocity, pressure, and wall-shear data was extracted. Velocity, pressure, friction factors, and form factors were retrieved from the corresponding Pronghorn cases for comparison. Figure~\ref{fig:velprofiles} shows the radial velocity profiles for each of the three Reynolds numbers. Similarly to the previous work, the KTA correlation was found to drastically overestimate the velocity in the outermost ring. The velocity in ring three was also moderately overestimated and the velocity in ring 4 was consistently underestimated. Form coefficients were manually altered to match the NekRS velocity results in the outermost, producing Pronghorn results that saw much better agreement with the NekRS case. A better understanding of why these changes cause the velocity profile to better reproduce the NekRS result can be obtained by investigating the friction and form coefficients between the two codes. Details on how the friction and form losses have been extracted from NekRS are provided in Section~\ref{sec:postpro}.

\begin{figure}
\centering
    \subfigure{
    \includegraphics[width=0.31\textwidth]{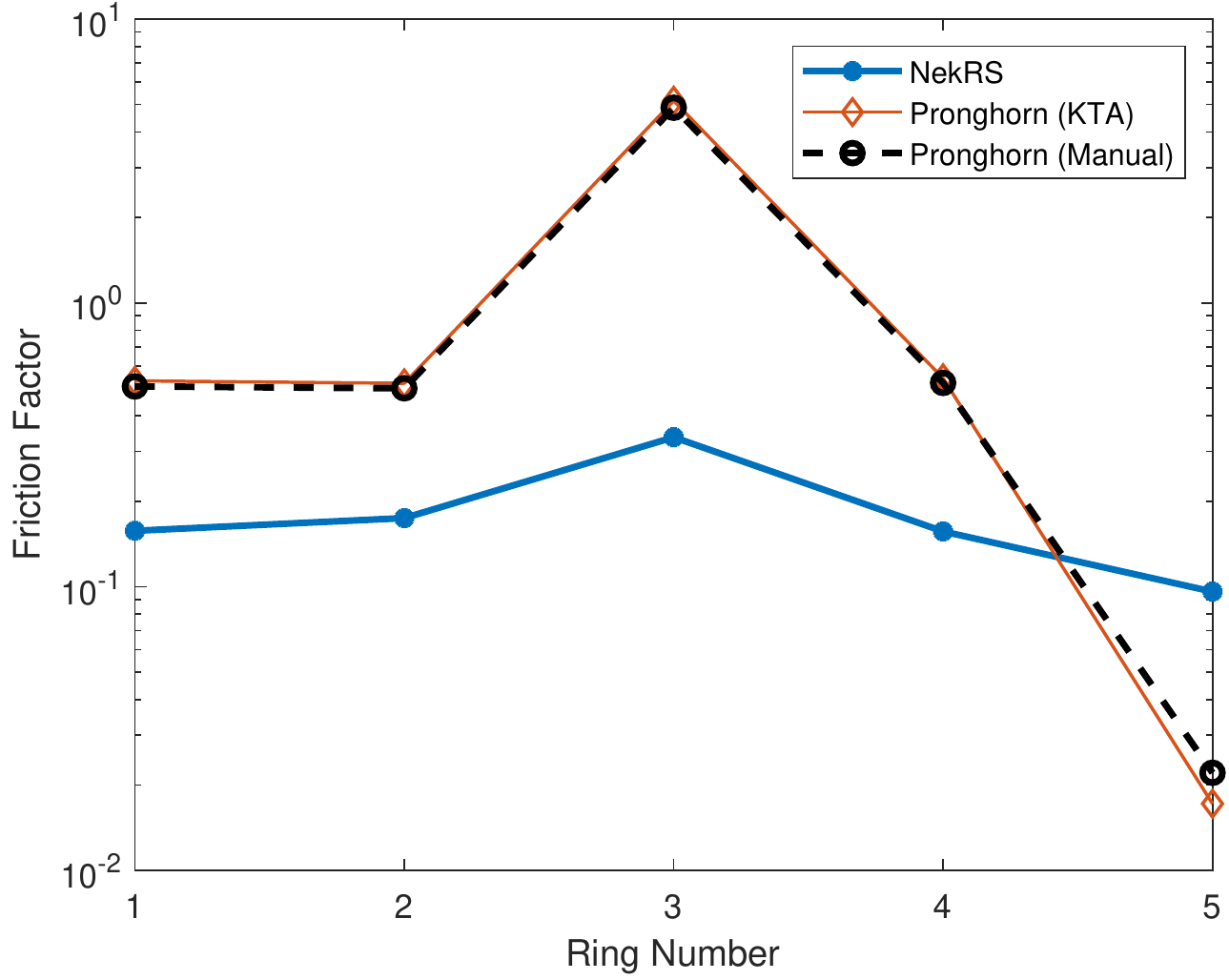}
    }
    \subfigure{
    \includegraphics[width=0.31\textwidth]{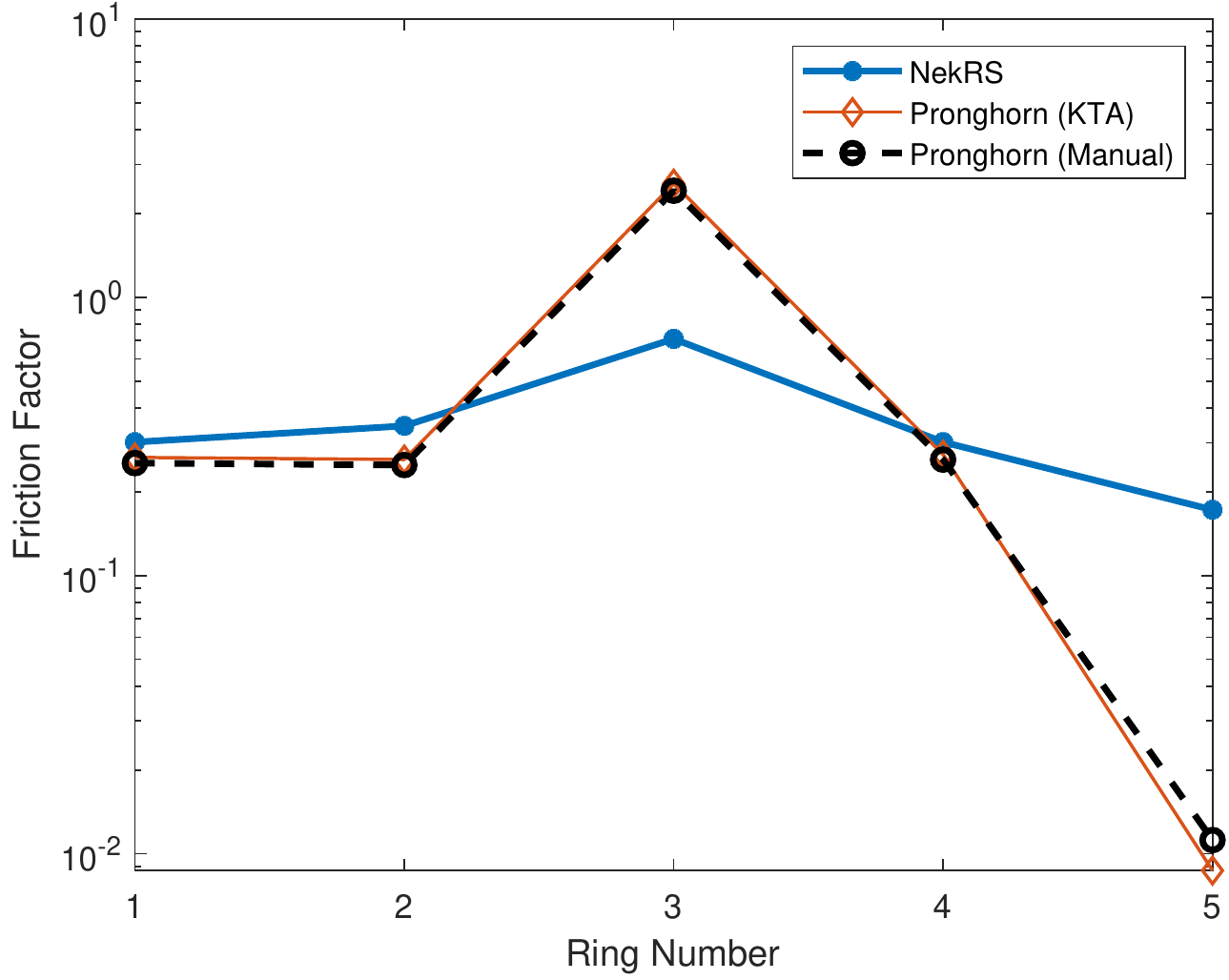}
    }
    \subfigure{
    \includegraphics[width=0.31\textwidth]{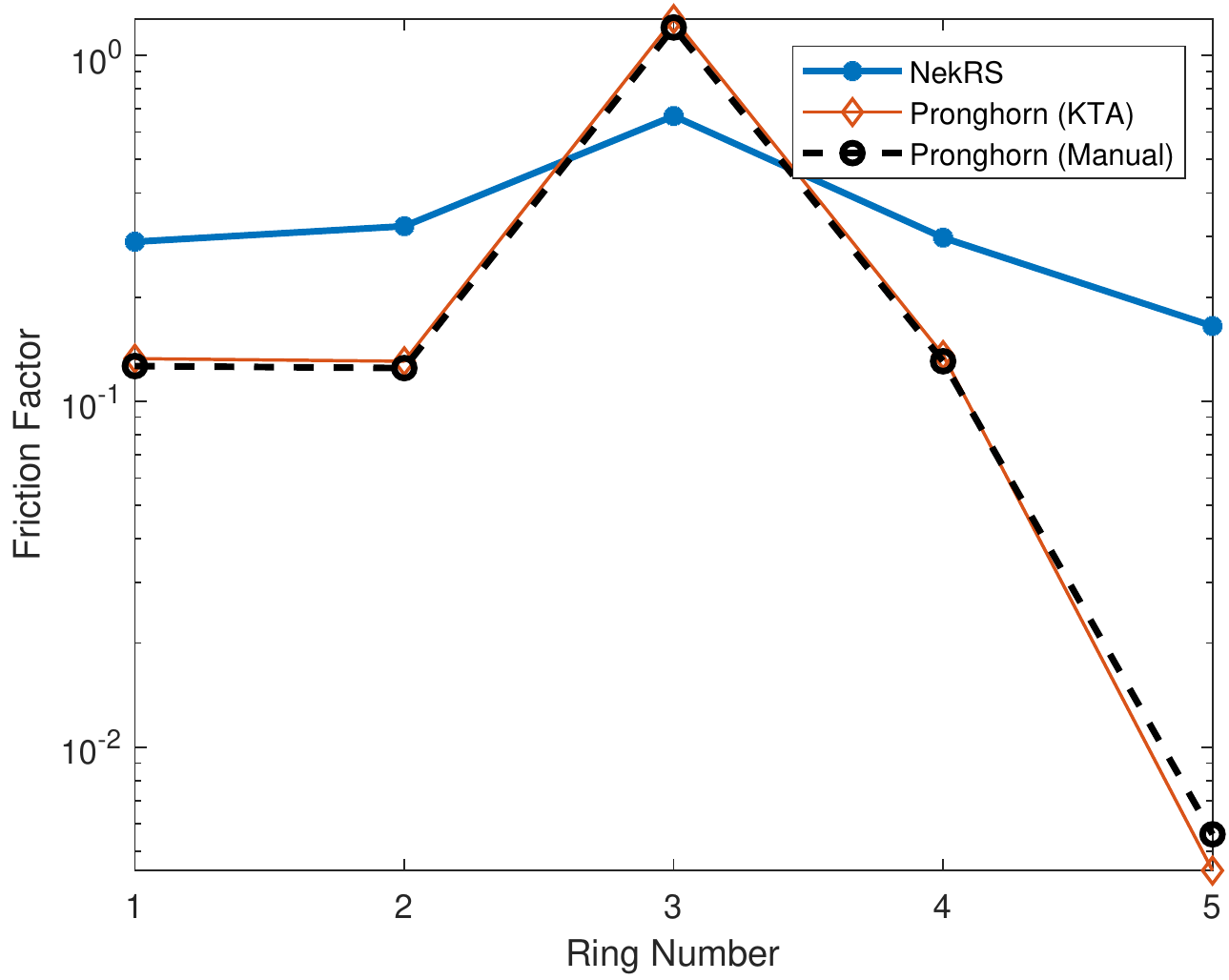}
    }
    
    \caption{Friction factor in each ring for the NekRS, Pronghorn KTA, and Pronghorn with manual adjustments cases for Re=2,500 (left), Re=5,000 (center) and Re=10,000 (right)}
    \label{fig:frictionfactors}
\end{figure}

\begin{figure}
\centering
    \subfigure{
    \includegraphics[width=0.31\textwidth]{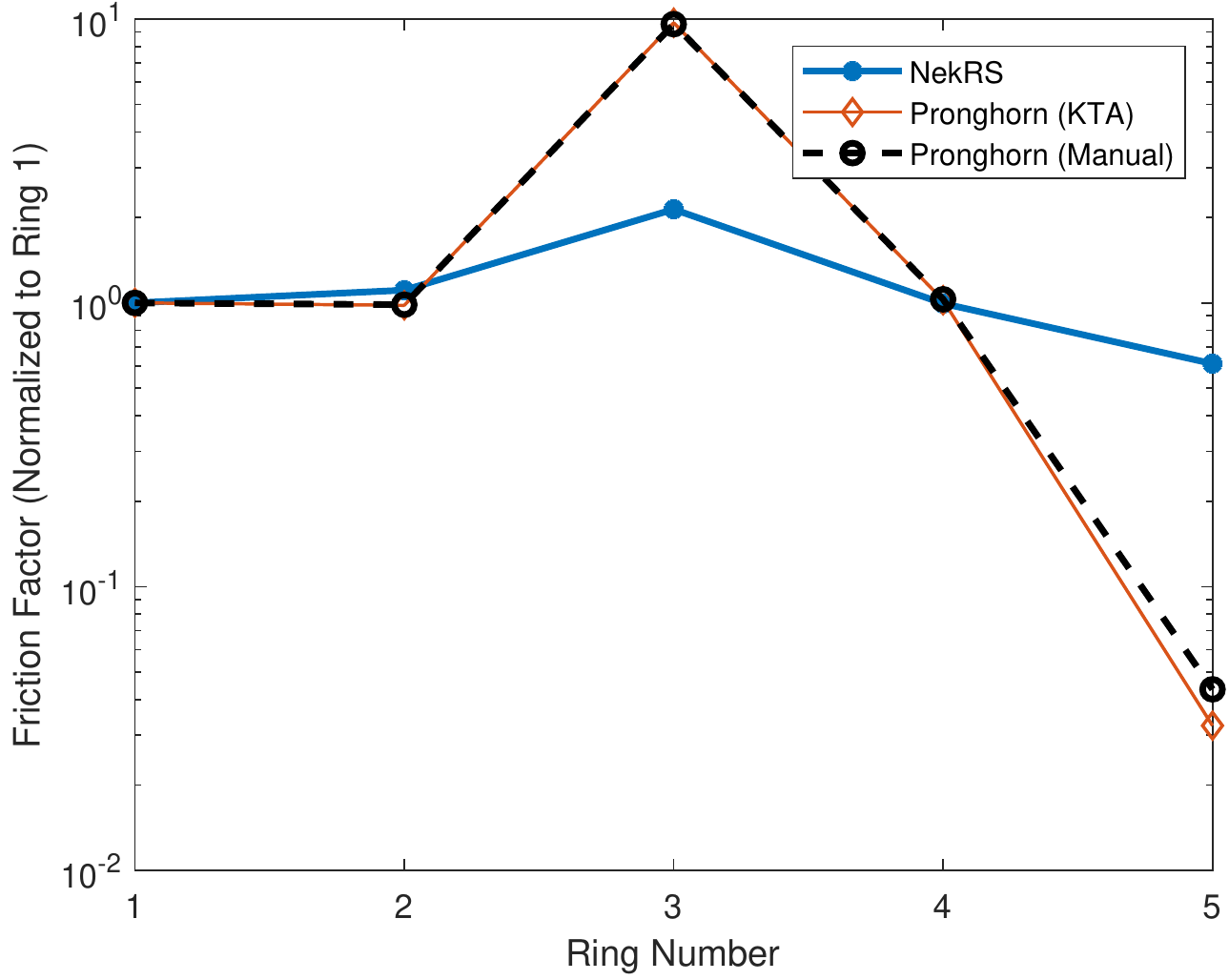}
    }
    \subfigure{
    \includegraphics[width=0.31\textwidth]{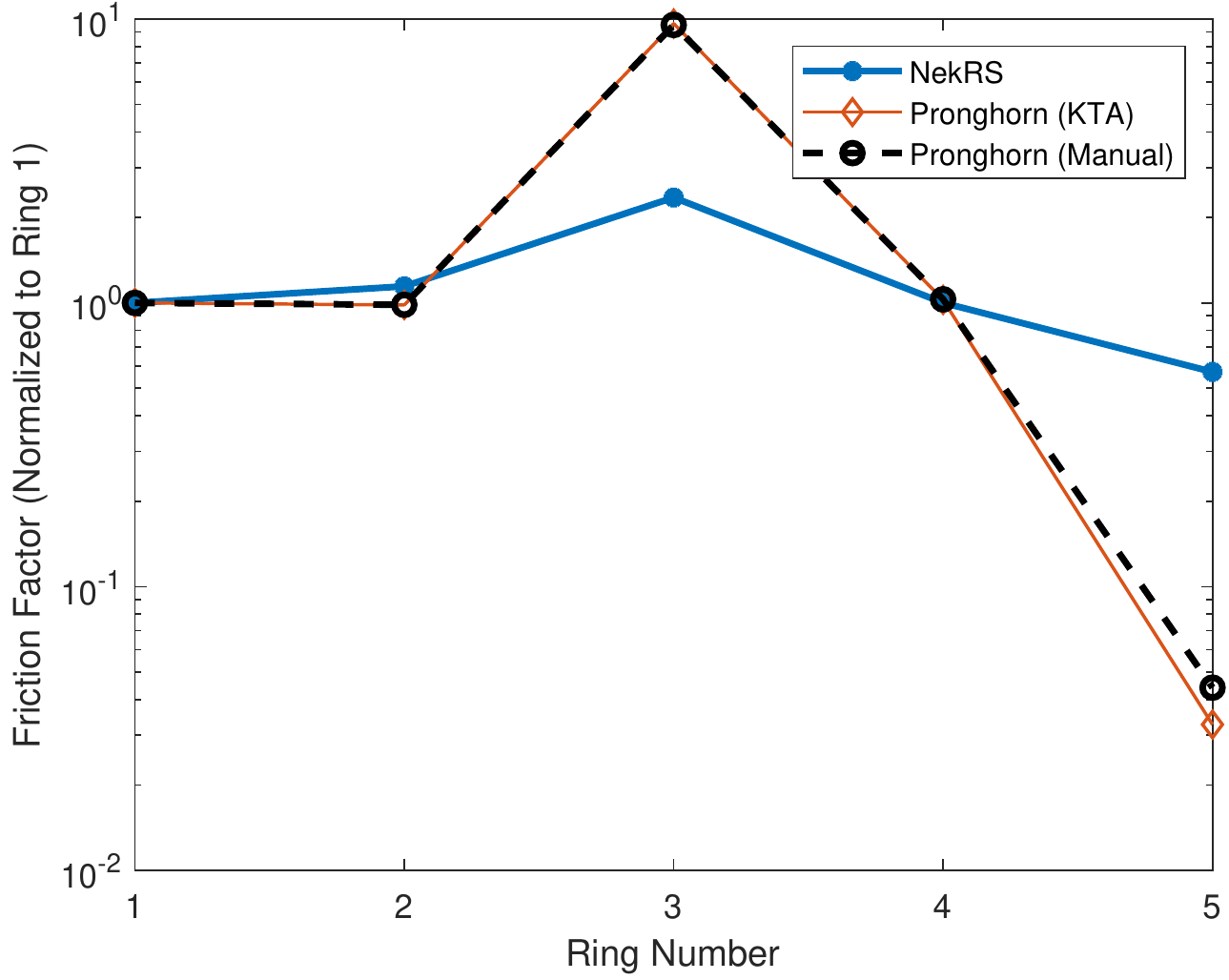}
    }
    \subfigure{
    \includegraphics[width=0.31\textwidth]{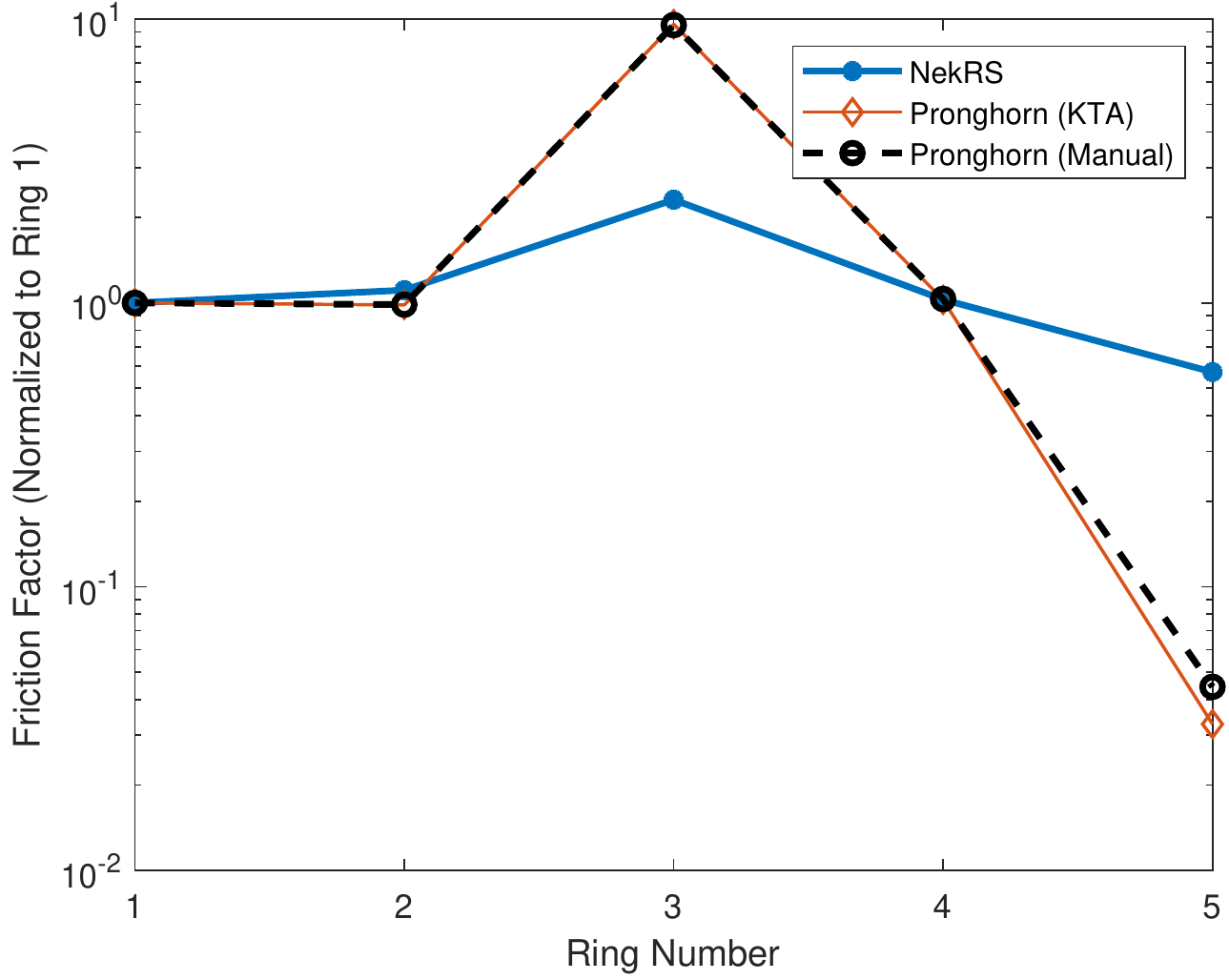}
    }
    
    \caption{Friction factor in each ring normalized by the first ring for the NekRS, Pronghorn KTA, and Pronghorn with manual adjustments cases for Re=2,500 (left), Re=5,000 (center) and Re=10,000 (right)}
    \label{fig:normfrictionfactors}
\end{figure}

The friction factor in each ring at each of the three Reynolds numbers is pictured in Figure~\ref{fig:frictionfactors}. The KTA correlation appears to overestimate the friction factor in all but the outermost ring for the Re = 2,500 case while a general underprediction of the friction factor is found for Re = 5,000 and Re = 10,000. Additionally, it is evident that the shape of the friction factor curve is much flatter in NekRS than what is determined by the KTA correlation. These values are also presented normalized by the first ring in Figure~\ref{fig:normfrictionfactors}. Although there is a significant disagreement between the NekRS and KTA results, this factor was not altered for this study due to the fact that the form losses (i.e., Forccheimer) are much more dominant in this case. Since only the form coefficients were altered in the manual Pronghorn case, there is only a slight difference in the friction factors compared to the KTA values. This change is a result of the velocity in each ring being different than when the KTA correlation is used for both friction and form. 

\begin{figure}[]
\centering
    \subfigure{
    \includegraphics[width=0.31\textwidth]{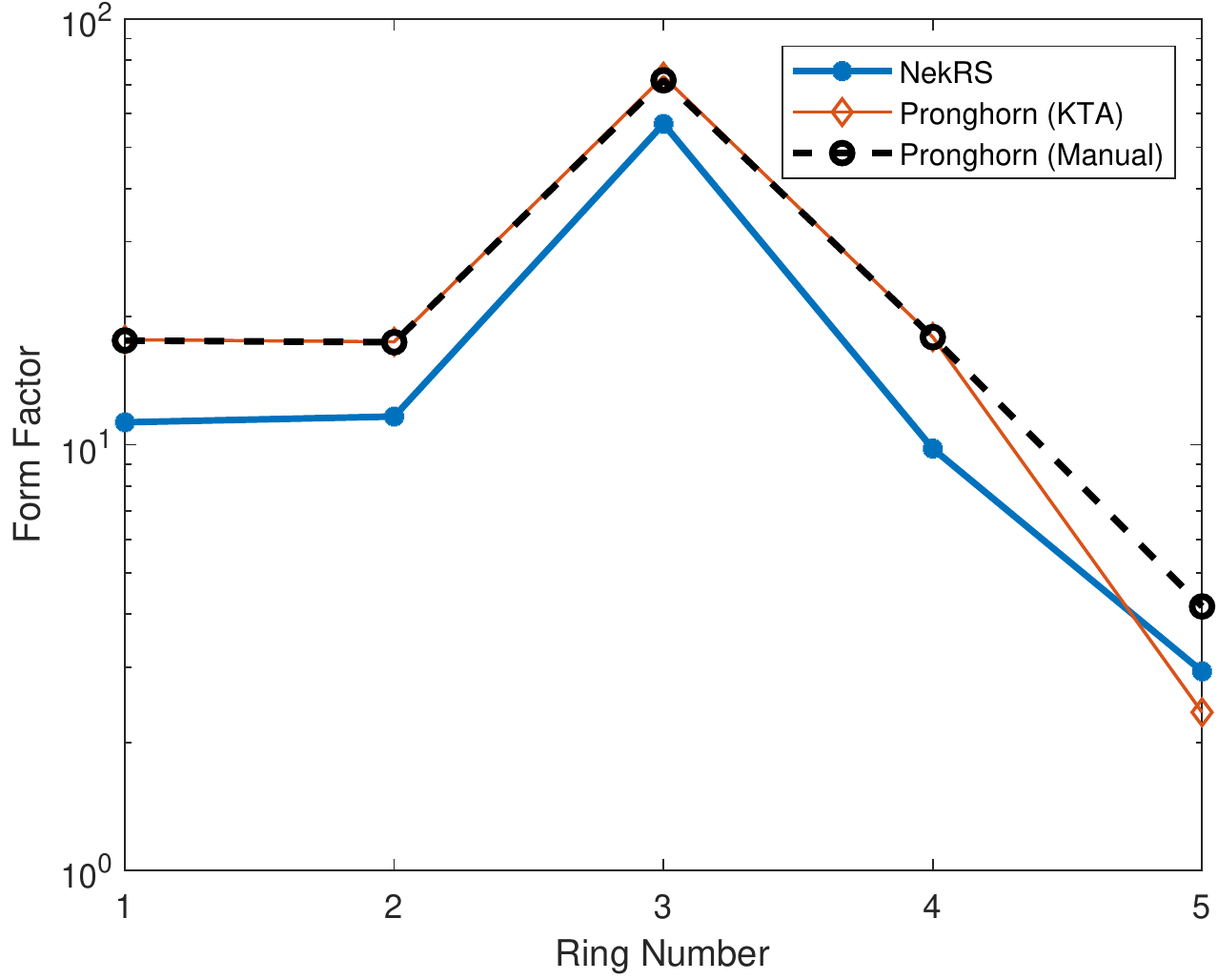}
    }
    \subfigure{
    \includegraphics[width=0.31\textwidth]{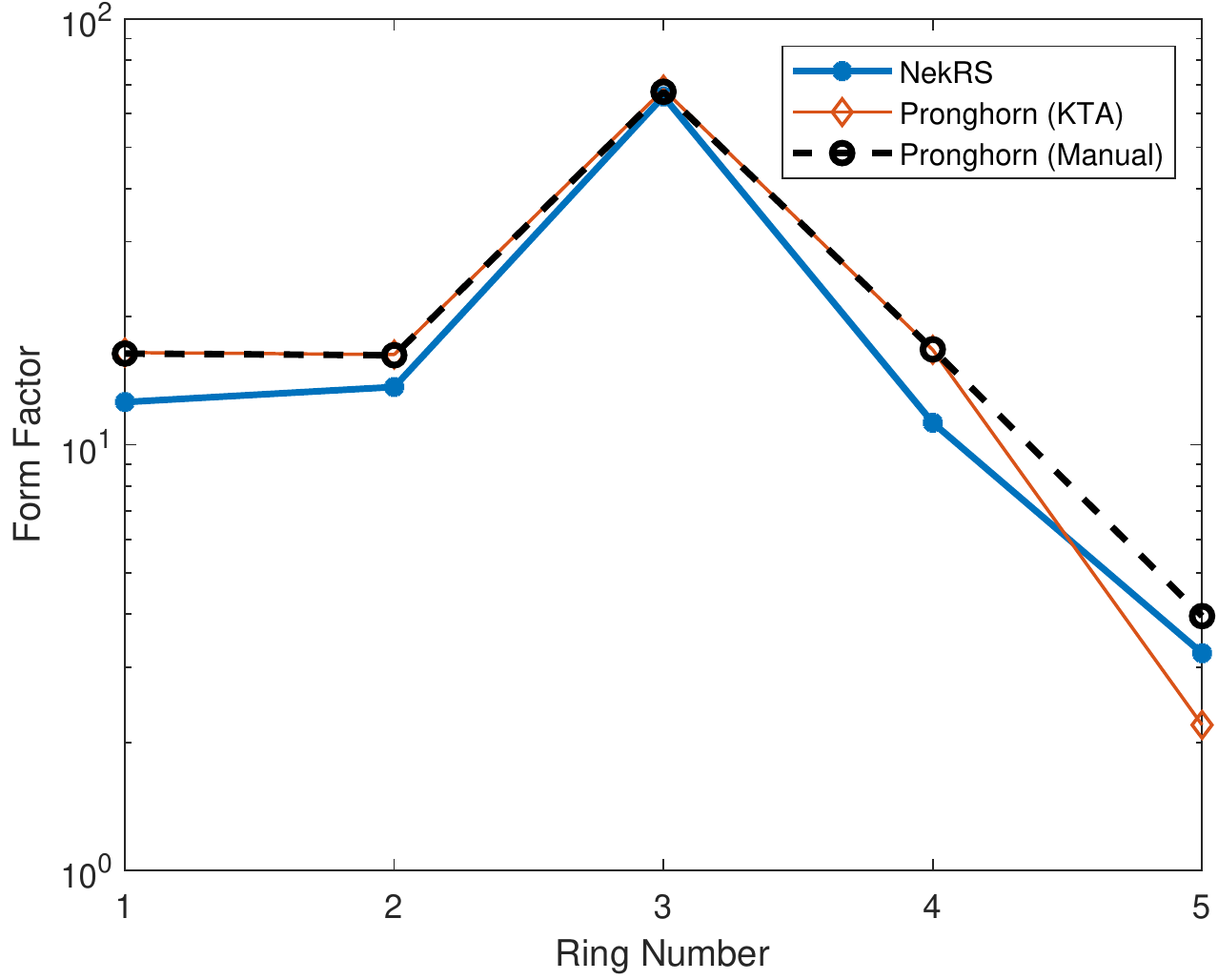}
    }
    \subfigure{
    \includegraphics[width=0.31\textwidth]{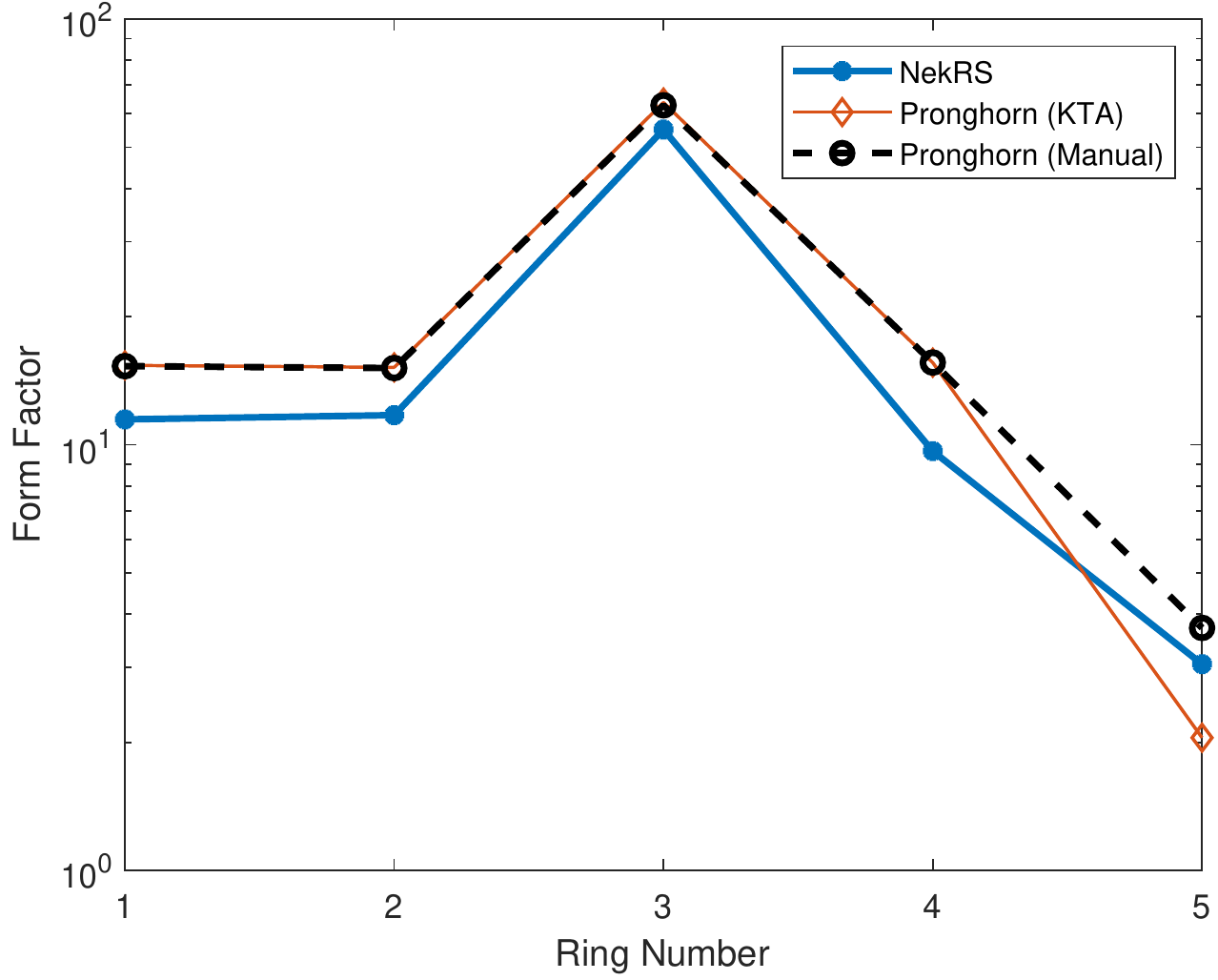}
    }
    
    \caption{Form factor in each ring for the NekRS, Pronghorn KTA, and Pronghorn with manual adjustments cases for Re=2,500 (left), Re=5,000 (center) and Re=10,000 (right)}
    \label{fig:formfactors}
\end{figure}

\begin{figure}
\centering
    \subfigure{
    \includegraphics[width=0.31\textwidth]{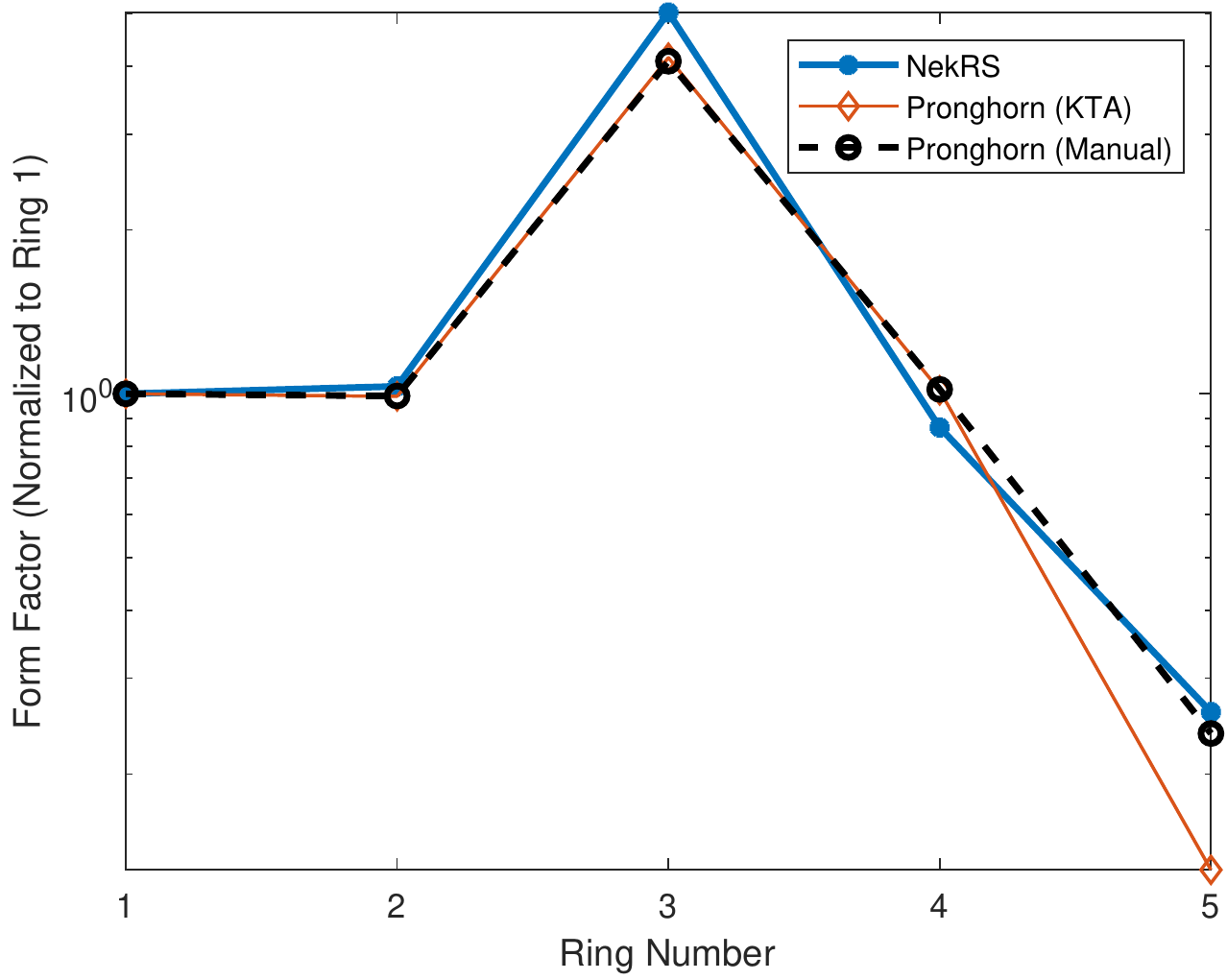}
    }
    \subfigure{
    \includegraphics[width=0.31\textwidth]{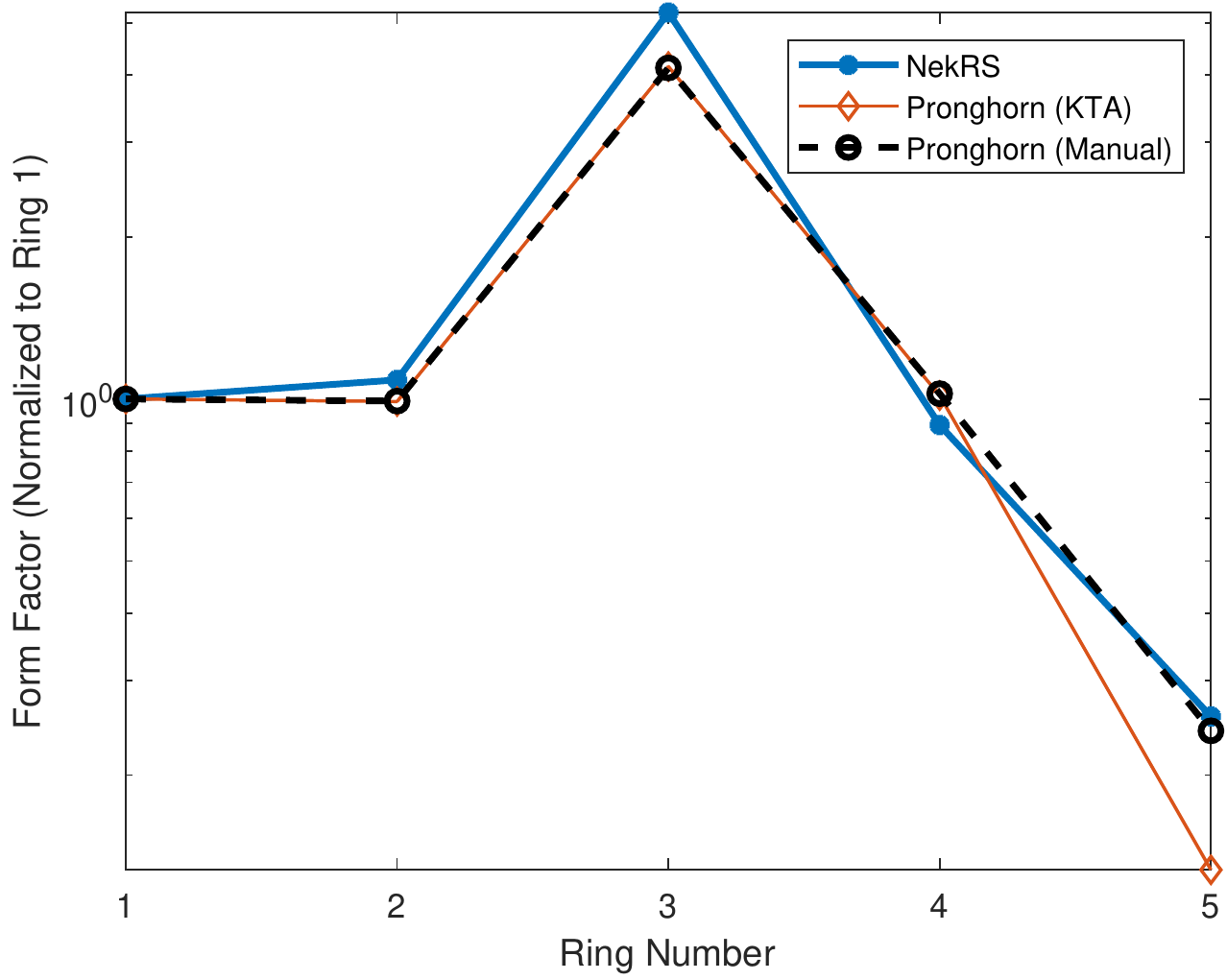}
    }
    \subfigure{
    \includegraphics[width=0.31\textwidth]{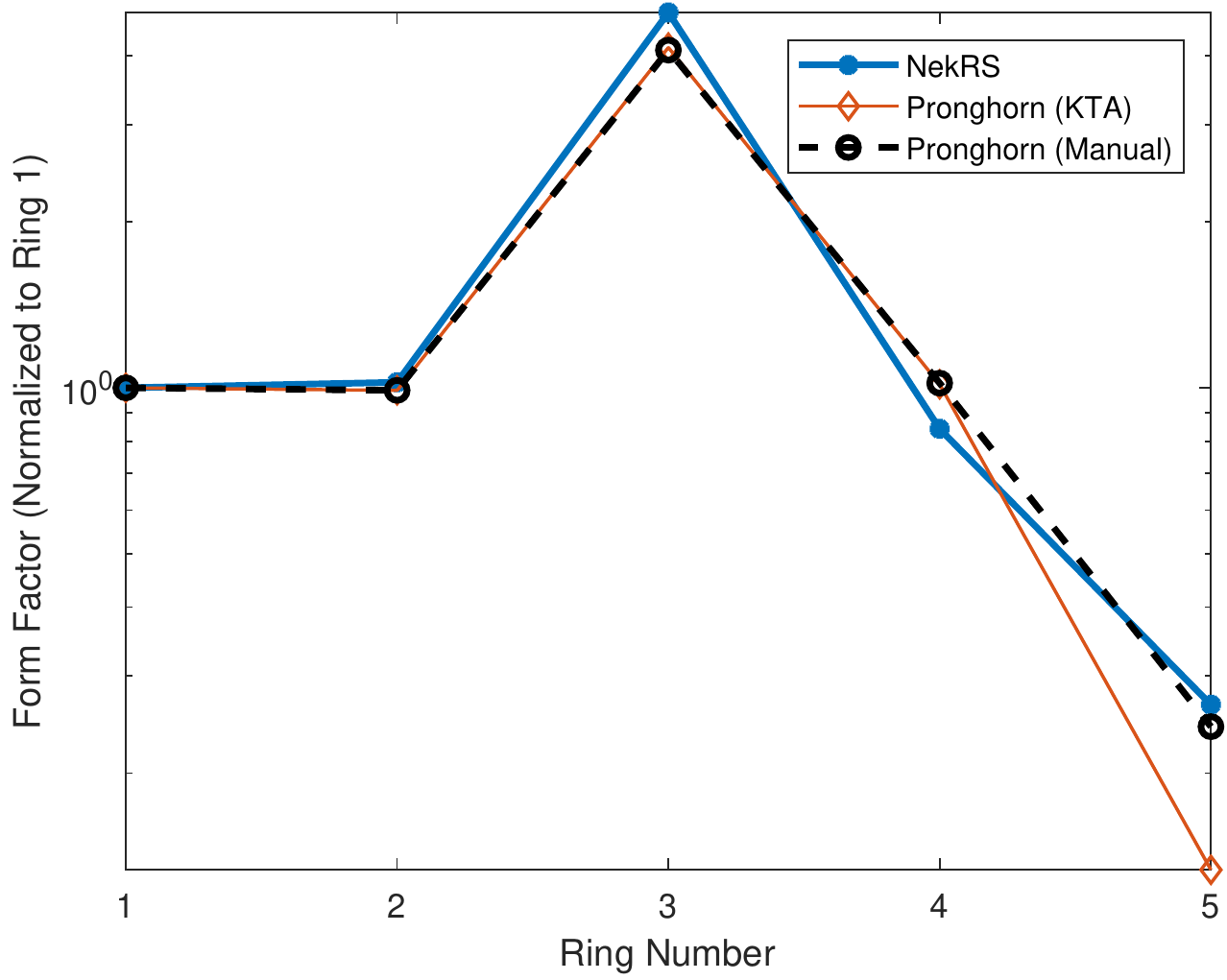}
    }
    
    \caption{Form factor in each ring normalized by the first ring for the NekRS, Pronghorn KTA, and Pronghorn with manual adjustments cases for Re=2,500 (left), Re=5,000 (center) and Re=10,000 (right)}
    \label{fig:normformfactors}
\end{figure}

Figure \ref{fig:formfactors} shows the form coefficients for each of the various cases. Much like the friction coefficient, the form coefficient was overestimated compared to the NekRS result in all but the outermost ring. This issue is worsened when looking at the manually adjusted coefficients, as the coefficients in ring five needed to be increased to match the NekRS velocity results. A better understanding of why the manual coefficients produce a velocity distribution in agreement with NekRS can be obtained by analysing the normalized form factors in Fig \ref{fig:normformfactors}. These results clearly display that the form factor in the near-wall region is much lower compared to the first ring with the KTA correlation than with NekRS. The manually adjusted Pronghorn values were adjusted such that the velocity in ring five matched the NekRS result, and the normalized form factor plots reveal that making these adjustments caused the normalized form factors in this region to match the NekRS result very well. This result is entirely expected and serves to further validate the methodology for producing a new correlation that may be used in Pronghorn to determine the correct coefficients a priori.

The pressure drop for the NekRS, KTA, and Wentz/Thodos cases can be found in Table \ref{tab:gradientcomparison}. The KTA correlation overpredicted the pressure drop at all three Reynolds numbers. Based on the previous analysis of the friction and form coefficients, this result is expected since the form coefficients were overpredicted in all but the outermost region. Likewise, the manually adjusted KTA case overpredicts the pressure drop even more since the form coefficients in ring  five needed to be increased to better match NekRS's velocity distribution. This result implies that although the form coefficient in the near-wall region needs to be increased with respect to the form coefficient in ring one, the coefficients in all rings need to be reduced if the pressure drop is to be matched between NekRS and the new correlation. The Wentz/Thodos correlation was also investigated due to its large range of porosity and Reynolds number validity. It performed, however, roughly similarly to the KTA correlation with around 7\% error compared to the NekRS result. The KTA correllation was much more consistent in its error, varying from 3-7\% error while the Wentz/Thodos correlation got siginificantly less accurate for higher Re, varying from around 1\% for the Re = 2,500 case to greater than 10\% in the Re =  10,000 case.

\begin{table}[]
    \centering
    \begin{tabular}{c|c|c|c}
        Source & Re = 2,500 & Re = 5,000 & Re = 10,000  \\ \hline
        NekRS & 7.46 & 6.87 & 6.37 \\
        KTA & 7.91 & 7.19 & 6.58 \\
        Wentz/Thodos & 7.59 & 6.50 & 5.67 \\
    \end{tabular}
    \caption{Comparison of pressure gradients predicted by NekRS, KTA correlation, and Wentz/Thodos correlation. Given pressure drops are nondimenionalized as $P_0/{d_{peb}}$ where $ P_0 = P/{\rho}{v}^2$}
    \label{tab:gradientcomparison}
\end{table}

\begin{figure}
    \centering
    \includegraphics[width=0.6\textwidth]{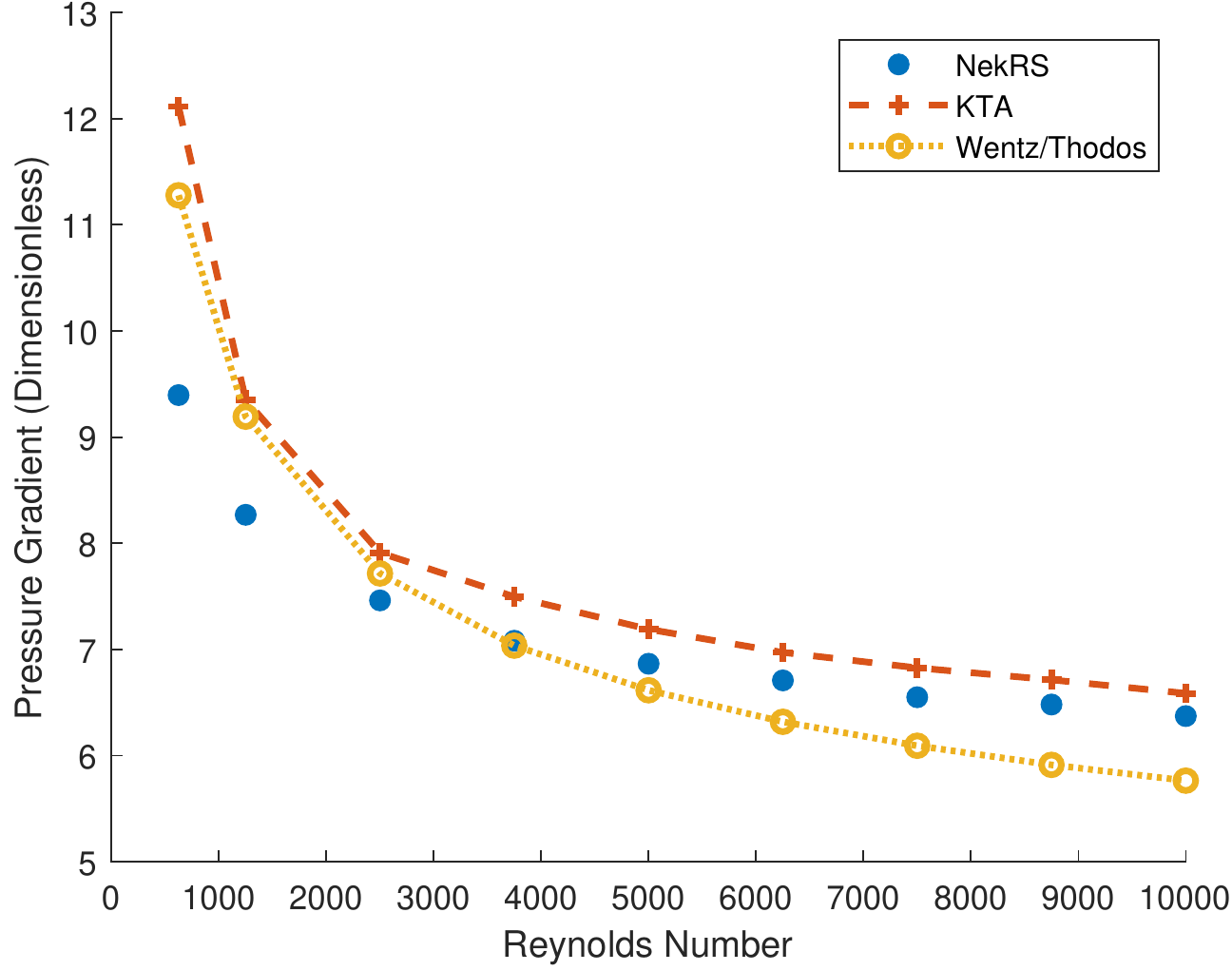}
    \caption{Dimensionless pressure gradient versus Reynolds number from NekRS, KTA correlation, and Wentz/Thodos correlation.}
    \label{fig:PressurePlot}
\end{figure}

After the initial analysis was performed on the three Reynolds numbers, data for additional Reynolds numbers was gathered ranging from Re = 625 to Re = 10,000. A comparison of the nondimensional pressure gradients determined with NekRS and from the two correlations is found in Figure~\ref{fig:PressurePlot}. The same trend experienced with the initial 3 Reynolds numbers is seen again, where the Wentz/Thodos correlation appears to have better agreement with the NekRS results for low Reynolds numbers, but deviates further at higher Reynolds numbers. The KTA correlation, meanwhile, is fairly consistent in its overprediction of the pressure gradient for Re $>$ 3,750.

\FloatBarrier
\section{Conclusions}

In this work, a pebble bed mesh was created with an average porosity of 0.44 and simulated using the spectral-element code NekRS. Cases were run for sufficiently long to produce a converged average flow field, and the average fields were extracted. Postprocessing was performed to determine the axial pressure profile, radial velocity profile, radial porosity profile, and radial wall shear profile of each of the three cases. With the pressure profiles known, the pressure gradients in the center of each case were determined and used for comparison to two correlations. The KTA correlation was chosen as the first comparative correlation due to its applicability to HTGR's. However, the porosity of the pebble mesh was slightly outside of the range of valid porosities for this correlation. The Wentz and Thodos correlation was chosen as the second comparative correlation, as it is said to have a wider valid porosity range that should be valid even in the near-wall region. The three Reynolds numbers used for this study were in the valid range of both correlations. 

An investigation of the pressure drop revealed that the KTA tended to overestimate the pressure drop with all three Reynolds numbers. The Wentz and Thodos correlation showed roughly equal agreement with the NekRS results at around 7\% error. It overpredicted the pressure drop for a Reynolds number of 2,500, but underpredicted the pressure drop for the Reynolds numbers of 5,000 and 10,000. Separating the pressure drop into the friction and form losses revealed that the pressure drop is largely dominated by form effects, with the form losses accounting for roughly 87\% of the total pressure drop in the inner rings and roughly 95\% in the near-wall region.

The velocity results from the two codes displayed an overprediction of the velocity in the near-wall region when using the KTA correlation in Pronghorn. The form coefficient in ring five was then manually increased so that the velocity in the near-wall region would match the results from the NekRS case. A closer investigation into the friction and form coefficients revealed the reason for this improved agreement. The friction factor from the KTA correlation was overpredicted compared to the NekRS result. It also was more sensitive to the porosity and Reynolds number of the ring, with the friction factor profile being much less flat than the profile from NekRS. The friction factor was not altered for this study since the form effects were shown to be dominant, so no significant change was observed when using the manually adjusted correlation. The form factor, like the friction factor, also displayed overprediction when using the KTA correlation. This overprediction explains why the KTA correlation produced a larger pressure gradient than what was seen in NekRS. When the form factor results were normalized about the innermost ring, it was evident that the form coefficient in ring five was lower than what was seen from the NekRS result. The manually adjusted value for ring five improved the velocity distribution. This in turn also leads to a slight worsening of the overprediction for the pressure drop since the form coefficient in ring five was increased even further. The normalized form factor plots, however, revealed that the adjusted values saw an improved agreement with the normalized NekRS results. This result aligns with intuition, as it is the magnitude of the coefficients that determines the pressure drop and the ratio between the coefficients in the rings that determines the velocity distribution.  With the process for this analysis now understood, it will be possible to determine improved coefficients for use in Pronghorn a priori, rather than manually adjusting them until the velocity results are more closely matched. 

Given that the KTA correlation is being used here slightly outside its range of validity, it is not surprising that it overpredicts the pressure drop. Coefficients would need to be decreased in magnitude across all regions to obtain better agreement with the NekRS result. Nonetheless, a crucial result of this work is that the form coefficient in the near-wall region needs to be increased with respect to the innermost bulk ring to prevent the overprediction of the velocity in this outer region. With the ability to determine form coefficients directly from the NekRS results, additional Reynolds number cases were simulated to generate more data to obtain a correlation for the form coefficient in the near-wall region. The data gathered from this study suggests that the the Forccheimer term of the KTA equation may altered in the near-wall region to the following:

\begin{equation}
    C_{Form} = \frac{8.4}{Re_m^{0.1}}
\end{equation}

\begin{figure}[h]
    \centering
    \includegraphics[width=0.6\textwidth]{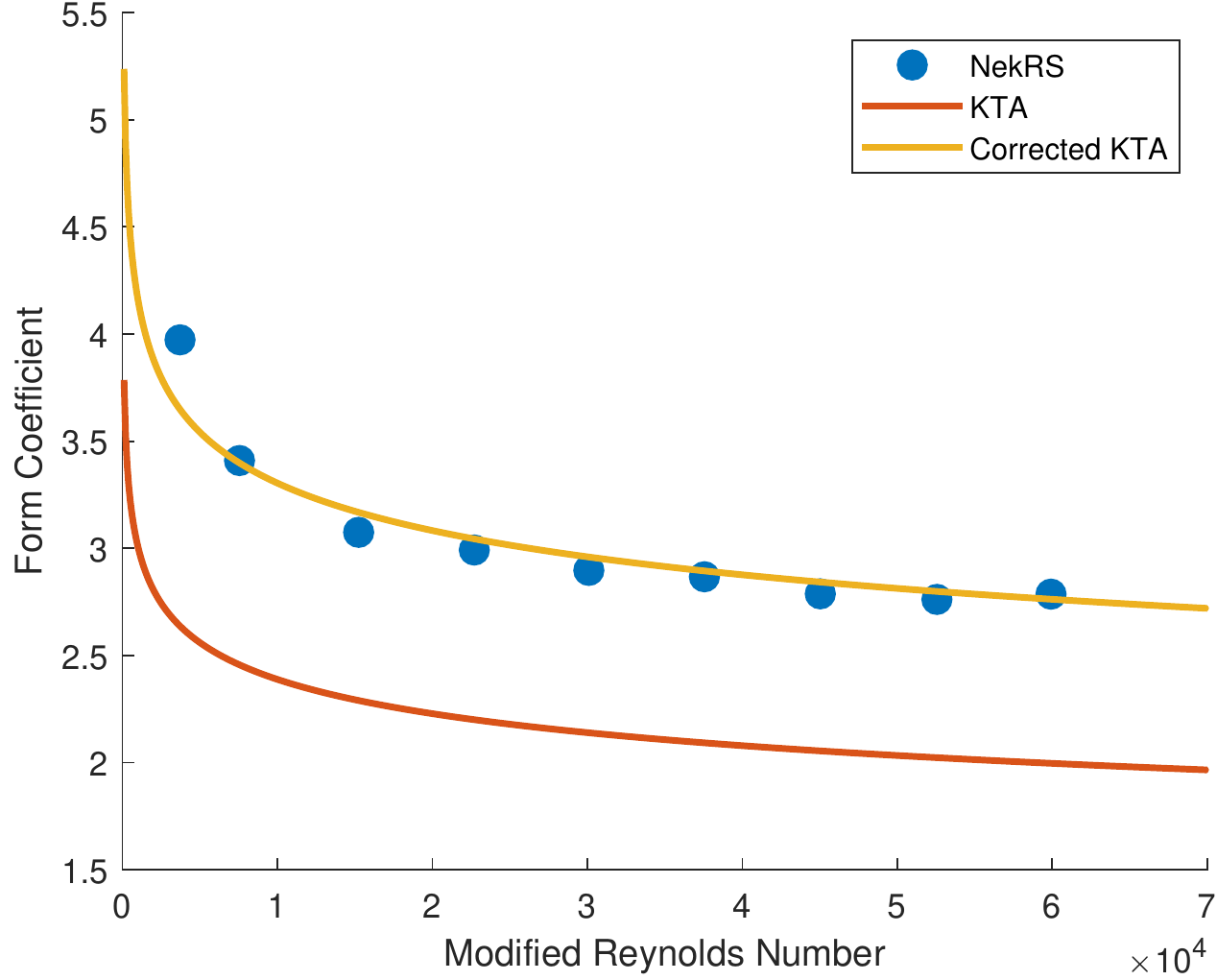}
    \caption{Forccheimer term in the near-wall region from NekRS, KTA, and the corrected KTA correlations.}
    \label{fig:NewForcc}
\end{figure}

\begin{figure}[h]
\centering
    \subfigure{
    \includegraphics[width=0.31\textwidth]{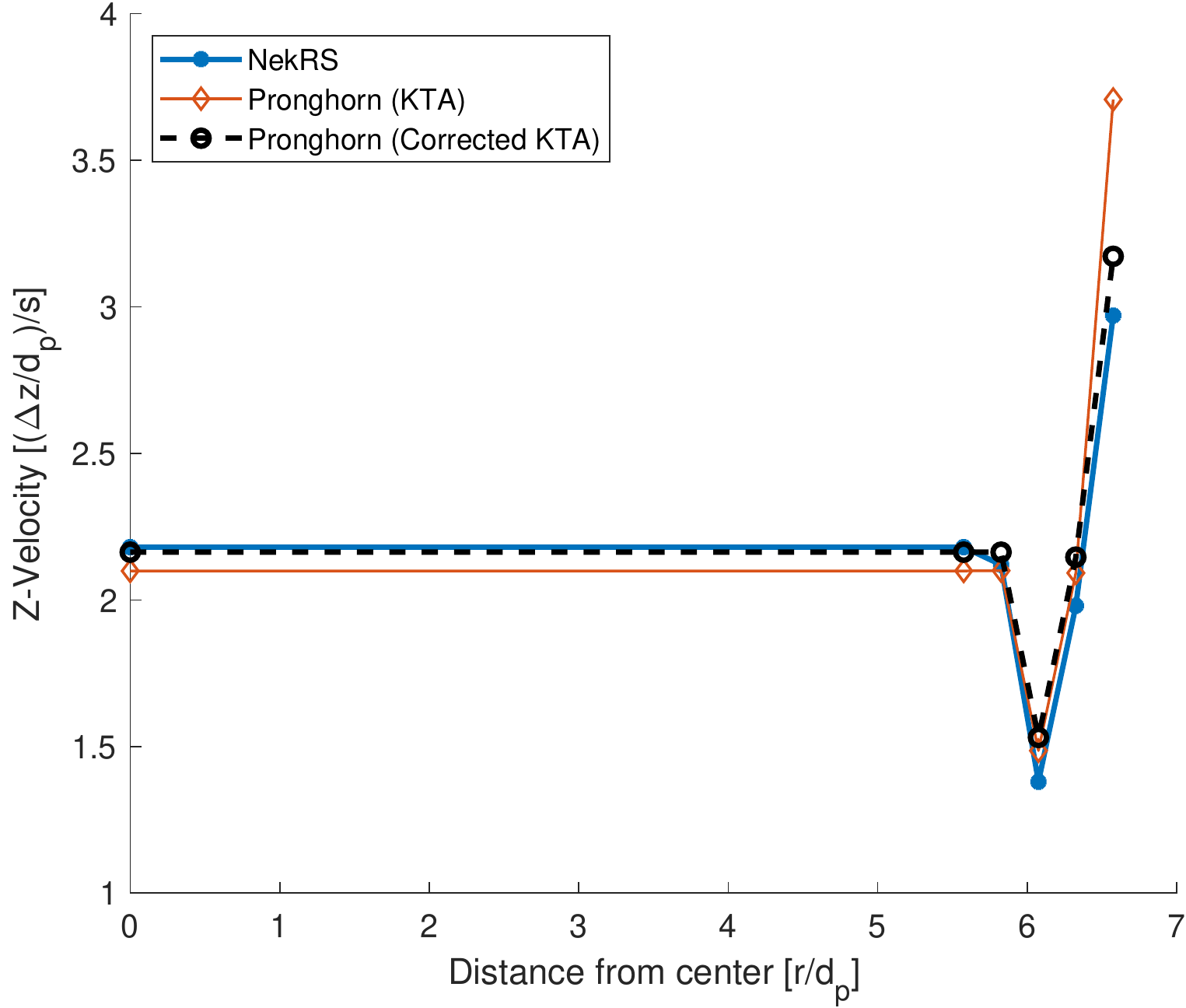}
    }
    \subfigure{
    \includegraphics[width=0.31\textwidth]{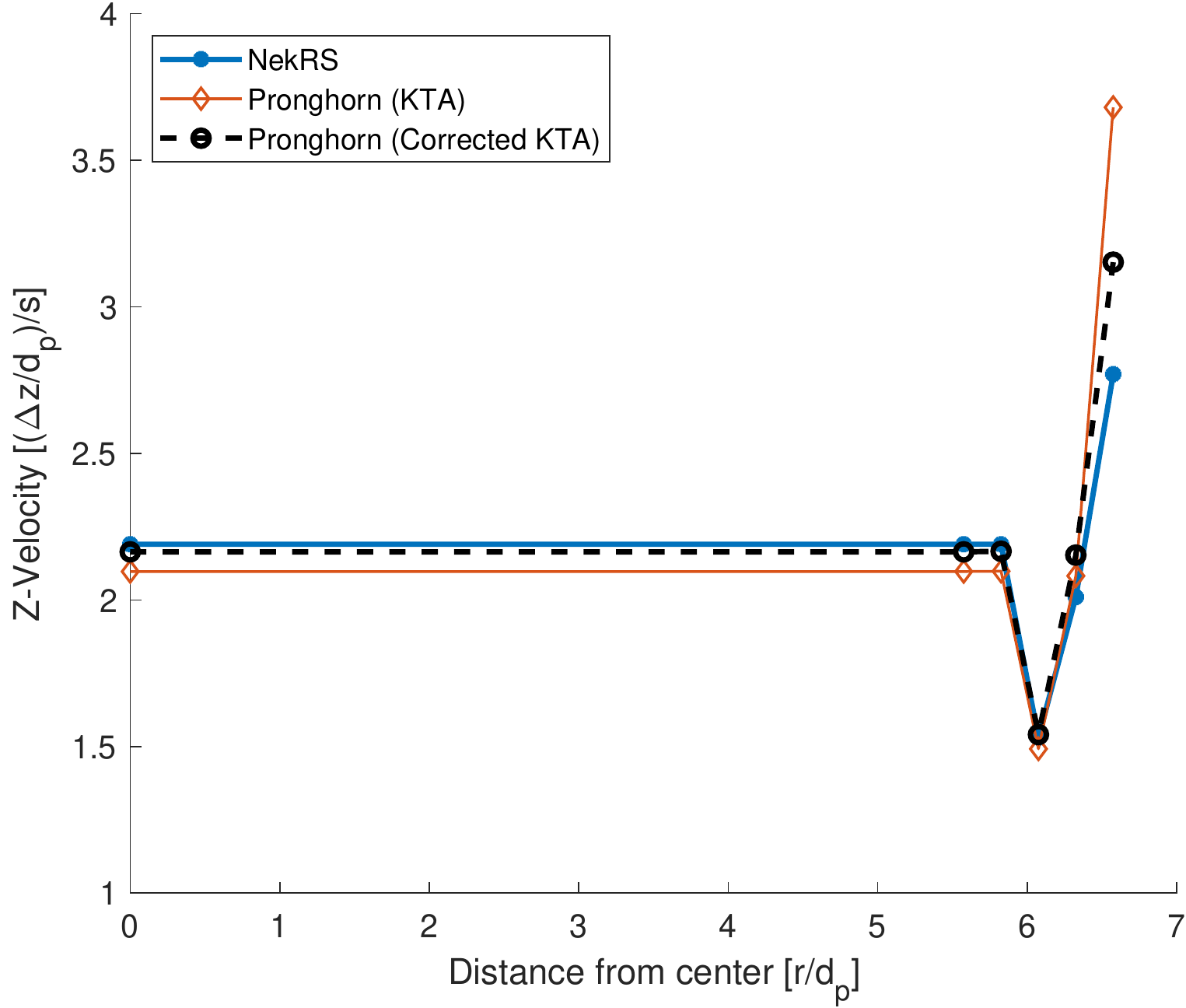}
    }
    \subfigure{
    \includegraphics[width=0.31\textwidth]{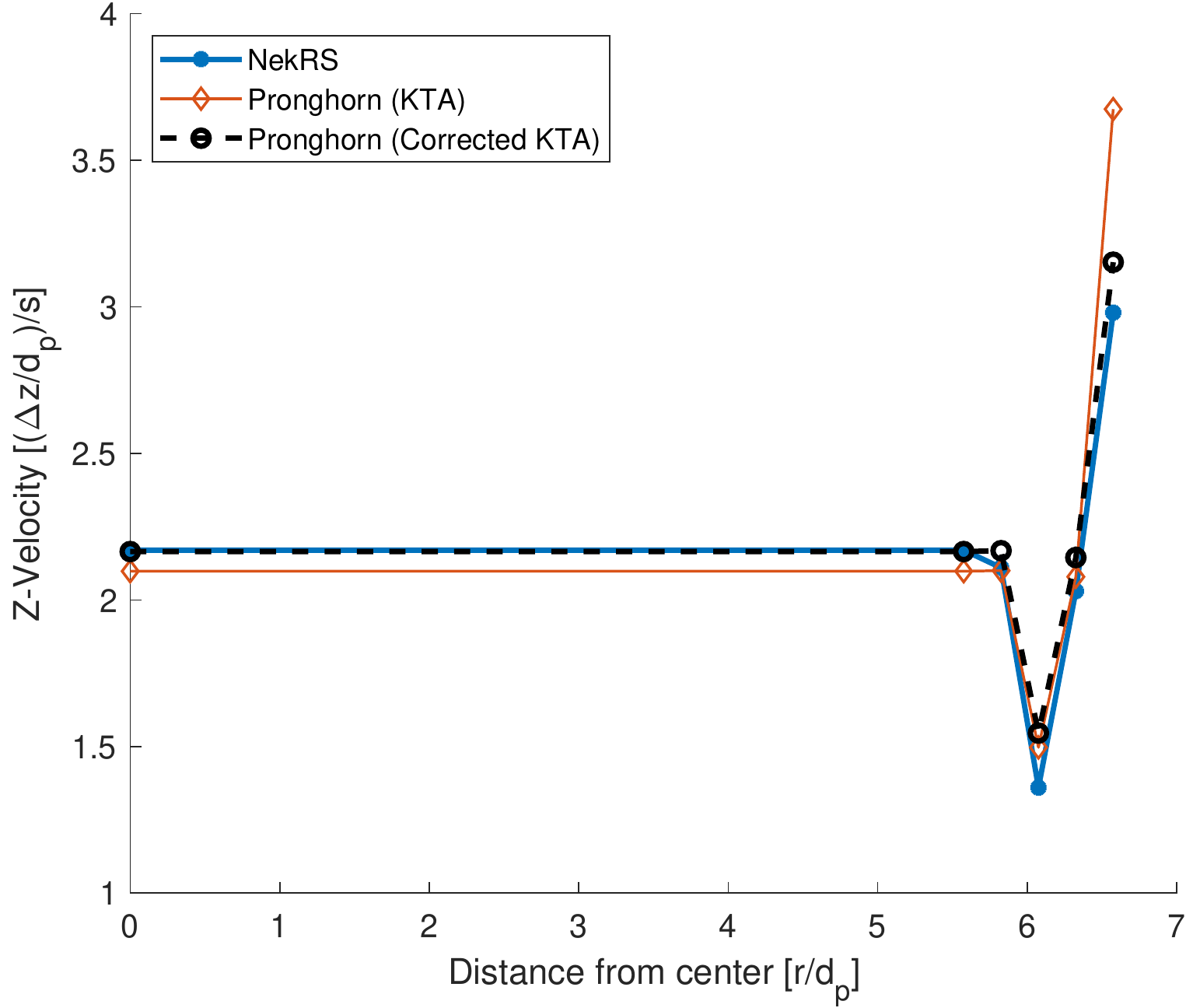}
    }
    
    \caption{Radial velocity profiles for the NekRS, Pronghorn KTA, and Pronghorn with corrected KTA correlation.}
    \label{fig:velprofilescorrected}
\end{figure}

The resulting form factors and velocity profiles when using this correction the outermost ring are pictured in Figures~\ref{fig:NewForcc} and \ref{fig:velprofilescorrected}.

The coefficients in the other four regions will still need to be reduced if the full NekRS velocity profile and pressure drop is to be matched. Future work will investigate the influence of the pebble packing ratio on the resulting form coefficient in the near-wall region, as this may significantly alter the agreement seen with the KTA correlation. Additionally, future work will aim to investigate and improve correlations for heat transfer in the near-wall region.
\FloatBarrier

\nomenclature[A]{CFD}{Computational Fluid Dynamics}
\nomenclature[A]{PBR}{Pebble Bed Reactor}
\nomenclature[A]{HTGR}{High-Temperature Gas-Cooled Reactor}
\nomenclature[A]{FHR}{Flouride-Cooled High-Temperature Reactor}
\nomenclature[A]{MOOSE}{Multiphysics Object-Oriented Simulation Environment}
\nomenclature[A]{DEM}{Discrete Element Method}
\nomenclature[A]{LES}{Large Eddy Simulation}
\nomenclature[A]{GPU}{Graphics Processing Unit}
\nomenclature[A]{DOE}{Department of Energy}
\nomenclature[A]{DNS}{Direct Numerical Simulation}
\nomenclature[v]{$D_h$}{Hydraulic Diameter}
\nomenclature[v]{$P$}{Pressure}
\nomenclature[v]{$\rho$}{Fluid Density}
\nomenclature[v]{$Re$}{Reynolds Number}
\nomenclature[v]{$Re_m$}{Modified Reynolds Number ($\frac{Re}{1-\epsilon})$}
\nomenclature[v]{$\tau_w$}{Wall Shear Stress}
\nomenclature[v]{$v_i$}{Interstitial Velocity}
\nomenclature[v]{$v_s$}{Superficial Velocity}
\nomenclature[v]{$\epsilon$}{Porosity}
\nomenclature[v]{$d_{peb}$}{Pebble Diameter}
\nomenclature[v]{$f$}{Darcy Friction Factor}
\nomenclature[v]{$C_{form}$}{Forccheimer Form Factor}
\printnomenclature

\bibliographystyle{unsrt}  
\bibliography{references}

\end{document}